\documentstyle[graphicx,lscape,times]{mn}

\newif\ifAMStwofonts
\ifoldfss
  \ifCUPmtlplainloaded \else
    \NewTextAlphabet{textbfit} {cmbxti10} {}
    \NewTextAlphabet{textbfss} {cmssbx10} {}
    \NewMathAlphabet{mathbfit} {cmbxti10} {} % for math mode
    \NewMathAlphabet{mathbfss} {cmssbx10} {} %  "   "    "
  \fi
  \ifAMStwofonts
    \ifCUPmtlplainloaded \else
      \NewSymbolFont{upmath} {eurm10}
      \NewSymbolFont{AMSa} {msam10}
      \NewMathSymbol{\upi}     {0}{upmath}{19}
      \NewMathSymbol{\umu}     {0}{upmath}{16}
      \NewMathSymbol{\upartial}{0}{upmath}{40}
      \NewMathSymbol{\leqslant}{3}{AMSa}{36}
      \NewMathSymbol{\geqslant}{3}{AMSa}{3E}

       \let\le=\leqslant
       
    \fi
  \fi
\fi % End of OFSS

\ifnfssone
  \newmathalphabet{\mathit}
  \addtoversion{normal}{\mathit}{cmr}{m}{it}
  \addtoversion{bold}{\mathit}{cmr}{bx}{it}
  \newmathalphabet{\mathbfit} % math mode version of \textbfit{..}
  \addtoversion{normal}{\mathbfit}{cmr}{bx}{it}
  \addtoversion{bold}{\mathbfit}{cmr}{bx}{it}
  \newmathalphabet{\mathbfss} % math mode version of \textbfss{..}
  \addtoversion{normal}{\mathbfss}{cmss}{bx}{n}
  \addtoversion{bold}{\mathbfss}{cmss}{bx}{n}
  \ifAMStwofonts
    \ifCUPmtlplainloaded \else
      \UseAMStwoboldmath
      \makeatletter
      \new@mathgroup\upmath@group
      \define@mathgroup\mv@normal\upmath@group{eur}{m}{n}
      \define@mathgroup\mv@bold\upmath@group{eur}{b}{n}
      \edef\UPM{\hexnumber\upmath@group}
      \new@mathgroup\amsa@group
      \define@mathgroup\mv@normal\amsa@group{msa}{m}{n}
      \define@mathgroup\mv@bold\amsa@group{msa}{m}{n}
      \edef\AMSa{\hexnumber\amsa@group}
      \makeatother
      \mathchardef\upi="0\UPM19
      \mathchardef\umu="0\UPM16
      \mathchardef\upartial="0\UPM40
      \mathchardef\leqslant="3\AMSa36
      \mathchardef\geqslant="3\AMSa3E

       \let\le=\leqslant

    \fi
  \fi
\fi % End of NFSS release 1

\ifnfsstwo
  \DeclareMathAlphabet{\mathbfit}{OT1}{cmr}{bx}{it}
  \SetMathAlphabet\mathbfit{bold}{OT1}{cmr}{bx}{it}
  \DeclareMathAlphabet{\mathbfss}{OT1}{cmss}{bx}{n}
  \SetMathAlphabet\mathbfss{bold}{OT1}{cmss}{bx}{n}
  \ifAMStwofonts
    \ifCUPmtlplainloaded \else
      \DeclareSymbolFont{UPM}{U}{eur}{m}{n}
      \SetSymbolFont{UPM}{bold}{U}{eur}{b}{n}
      \DeclareSymbolFont{AMSa}{U}{msa}{m}{n}
      \DeclareMathSymbol{\upi}{0}{UPM}{"19}
      \DeclareMathSymbol{\umu}{0}{UPM}{"16}
      \DeclareMathSymbol{\upartial}{0}{UPM}{"40}
      \DeclareMathSymbol{\leqslant}{3}{AMSa}{"36}
      \DeclareMathSymbol{\geqslant}{3}{AMSa}{"3E}

       \let\le=\leqslant

    \fi
  \fi
\fi % End of NFSS release 2

\ifCUPmtlplainloaded \else
  \ifAMStwofonts \else % If no AMS fonts
    \def\upi{\pi}
    \def\umu{\mu}
    \def\upartial{\partial}
  \fi
\fi

\title[ACS photometry of old LMC clusters]
  {Photometry of Magellanic Cloud clusters with the Advanced Camera for Surveys - I. The old LMC clusters NGC 1928, 1939 and Reticulum}
\author[A.~D.~Mackey \& G.~F.~Gilmore]
  {A.~D.~Mackey$^1$\thanks{E-mail: dmackey@ast.cam.ac.uk}
  and G.~F.~Gilmore$^1$\\
  $^1$Institute of Astronomy, University of Cambridge, Madingley Road,
  Cambridge CB3 0HA}
\date{Accepted --. Received --}
\pagerange{000--000}
\pubyear{2004}

\def\LaTeX{L\kern-.36em\raise.3ex\hbox{a}\kern-.15em
    T\kern-.1667em\lower.7ex\hbox{E}\kern-.125emX}

\begin{document}

\label{firstpage}

\maketitle

\begin{abstract}
We present the results of photometric measurements from images of the LMC globular clusters
NGC 1928, 1939 and Reticulum taken with the Advanced Camera for Surveys on the 
{\em Hubble Space Telescope}. Exposures through the F555W and F814W filters result in high accuracy
colour-magnitude diagrams (CMDs) for these three clusters. This is the first time that CMDs
for NGC 1928 and 1939 have been published. All three clusters possess CMDs with
features indicating them to be $> 10$ Gyr old, including  main sequence turn-offs at 
$V\sim 23$ and well populated horizontal branches (HBs). We use the CMDs to obtain metallicity
and reddening estimates for each cluster. NGC 1939 is a metal-poor cluster, with 
$[{\rm Fe}/{\rm H}] = -2.10 \pm 0.19$, while NGC 1928 is significantly more metal-rich, with 
$[{\rm Fe}/{\rm H}] = -1.27 \pm 0.14$. The abundance of Reticulum is intermediate between the two, 
with $[{\rm Fe}/{\rm H}] = -1.66 \pm 0.12$ -- a measurement which matches well with previous 
estimates. All three clusters are moderately reddened, with values ranging from 
$E(V-I) = 0.07 \pm 0.02$ for Reticulum and $E(V-I) = 0.08 \pm 0.02$ for NGC 1928, to 
$E(V-I) = 0.16 \pm 0.03$ for NGC 1939. After correcting the CMDs for extinction we estimate the HB 
morphology of each cluster. NGC 1928 and 1939 possess HBs consisting almost exclusively of stars to 
the blue of the instability strip, with NGC 1928 in addition showing evidence for an extended blue 
HB. In contrast, Reticulum has an intermediate HB morphology, with stars across the instability 
strip. Using a variety of dating techniques we show that these three clusters are coeval with each 
other and the oldest Galactic and LMC globular clusters, to within $\sim 2$ Gyr. The census of 
known old  LMC globular clusters therefore now numbers $15$ plus the unique, somewhat younger 
cluster ESO121-SC03. The NGC 1939 field contains another cluster in the line-of-sight, NGC 1938. A 
CMD for this object shows it to be less than $\sim 400$ Myr old, and it is therefore unlikely to be 
physically associated with NGC 1939.
\end{abstract}


\section{Introduction}
\label{s:intro}
The Large and Small Magellanic Clouds (LMC/SMC) possess extensive systems of rich stellar clusters.
These objects exhibit a much wider variety in age, structure, environment and mass than do Galactic
clusters, and this, combined with their relatively close proximity, has rendered them central
to a surprising number of fields of modern astrophysics -- from star and cluster formation, and
stellar evolution, to gravitational dynamics, galactic evolution, and distance scale measurements. 
They are also vital probes and tracers of the chemical and dynamical evolution of the LMC and SMC 
themselves. It is therefore important to understand how their properties, and in particular their 
ages and abundances, are distributed.

It has long been known that the LMC, which contains the more numerous cluster system of the two
Clouds, houses a small sub-population of extremely ancient objects. Although early studies
revealed half a dozen LMC clusters to possess colour-magnitude diagrams (CMDs) with features
similar to those of the Galactic globular clusters, it has only been since the advent of the 
{\em Hubble Space Telescope} ({\em HST}) that imaging resolution and sensitivity has been 
sufficiently high as to allow accurate colour-magnitude diagrams (CMDs) suitable for relative age 
dating. A number of relatively recent studies have demonstrated that the number of LMC clusters
coeval with each other and the oldest Galactic globular clusters is somewhat more than a dozen --
NGC 1466, NGC 2257, and Hodge 11 (e.g., Johnson et al. \shortcite{johnson:99}); NGC 1754, 1835,
1898, 1916, 2005, and 2019 (e.g., Olsen et al. \shortcite{olsen}); and NGC 1786, 1841 and 2210
(e.g., Brocato et al. \shortcite{brocato}). Several published CMDs show the remote outer
cluster Reticulum to be very old \cite{johnson:99,marconi,monelli}; however a full age analysis
is yet to be published for this cluster. In addition there are two more clusters located in the
LMC bar region -- NGC 1928 and 1939 -- which integrated spectroscopy suggests could be old 
\cite{dutra}, but which are so compact and lie against such highly crowded LMC fields that it has 
not previously been possible to obtain accurate CMDs for them (see for example, the search for old 
LMC clusters conducted by Geisler et al. \shortcite{geisler}). The census of old LMC clusters is 
therefore still incomplete.

We have used the Advanced Camera for Surveys (ACS) on {\em HST} to obtain images of 
NGC 1928, 1939, and Reticulum as part of program 9891 -- a snapshot survey of 
$40$ LMC and $40$ SMC clusters. This program is primarily designed to extend the detailed 
investigation of Mackey \& Gilmore \shortcite{sbp1,sbp2} concerning the structural evolution of 
Magellanic Cloud clusters. However, the ACS observations are of sufficient quality and resolution
as to allow colour-magnitude diagrams (CMDs) to be constructed for NGC 1928 and NGC 1939 for the 
first time, as well as a high quality CMD for Reticulum. In this paper, we present the results of 
these observations (Section \ref{s:obsred}) and the measured colour-magnitude diagrams (Section
\ref{s:colourmag}). Abundances and reddenings are derived for each cluster, and it is demonstrated 
that NGC 1928 and 1939, along with Reticulum, are coeval with both well studied Galactic globular 
clusters (such as M92, M3 and M5) and other old LMC objects (such as NGC 2257 and Hodge 11)
(Section \ref{s:clusterprop}). 

\section{Observations and Data Reduction}
\label{s:obsred}
The observations were taken during {\em HST} Cycle 12 using the ACS Wide Field Channel (WFC). 
As snapshot targets, the clusters were observed for only one orbit each. This allowed two 
exposures to be taken per cluster -- one through the F555W filter and one through the F814W
filter. Details of the individual exposures are listed in Table \ref{t:observations}. Exposure
durations were $330$ s in F555W and $200$ s in F814W. 

The ACS WFC consists of a mosaic of two $2048 \times 4096$ SITe CCDs with a scale of
$\sim 0.05$ arcsec per pixel, and separated by a gap of $\sim 50$ pixels. Each image therefore 
covers a field of view (FOV) of approximately $202 \times 202$ arcseconds. The clusters were 
centred at the reference point WFC1, located on chip 1 at position $(2072\,,\,1024)$. This 
allowed any given cluster to be observed up to a radius $r\sim 150 \arcsec$ from its centre,
while also ensuring the cluster core did not fall near the inter-chip gap.

In order to maximise the efficiency of the limited imaging strategy afforded by a snapshot 
program, all observations were made with the ACS/WFC GAIN parameter set to $2$ rather than the
default (GAIN $=1$). This allowed the full well depth to be sampled (as opposed to only 
$\sim 75$ per cent of the full well depth for GAIN $=1$) with only a modest increase in read noise
($\sim 0.3$ e$^{-}$ extra rms), thus increasing the dynamic range of the observations by
greater than $0.3$ mag. In addition, the second image of each cluster was offset by $2$ pixels
in both the $x$ and $y$ directions, to help facilitate the removal of hot pixels and cosmic 
rays. With only two images per cluster, through different filters, it is not possible to completely
eliminate the inter-chip gap using such an offset.

\begin{table*}
\begin{minipage}{155mm}
\caption{ACS/WFC observations of NGC 1928, 1939, and Reticulum ({\em HST} program 9891).}
\begin{tabular}{@{}lcccccccc}
\hline \hline
Cluster & \hspace{10mm} & RA & Dec. & \hspace{10mm} & Filter & Dataset & Exposure & Date \\
 & & (J2000.0) & (J2000.0) & & & & Time (s) & \\
\hline
NGC 1928 & & $05^{{\rm h}}\,\,20^{{\rm m}}\,\,57.51^{{\rm s}}$ & $-69\degr\,\,28\arcmin\,\,41.5\arcsec$ & & F555W & j8ne62ztq & $330$ & $23/08/2003$ \\
 & & & & & F814W & j8ne62zvq & $200$ & $23/08/2003$ \\
NGC 1939 & & $05^{{\rm h}}\,\,21^{{\rm m}}\,\,26.63^{{\rm s}}$ & $-69\degr\,\,56\arcmin\,\,58.2\arcsec$ & & F555W & j8ne63tqq & $330$ & $27/07/2003$ \\
 & & & & & F814W & j8ne63ttq & $200$ & $27/07/2003$ \\
Reticulum & & $04^{{\rm h}}\,\,36^{{\rm m}}\,\,09.33^{{\rm s}}$ & $-58\degr\,\,51\arcmin\,\,40.3\arcsec$ & & F555W & j8ne43a3q & $330$ & $21/09/2003$ \\
 & & & & & F814W & j8ne43a7q & $200$ & $21/09/2003$ \\
\hline
\label{t:observations}
\end{tabular}
\end{minipage}
\end{table*}

Before being made available for retrieval from the STScI archive, all images were passed through
the standard ACS/WFC reduction pipeline. This process includes bias and dark subtractions,
flatfield division, masking of known bad pixels and columns, and the calculation of photometry 
header keywords. In addition the STScI {\em PyDrizzle} software is used to correct the 
(significant) geometric distortion present in WFC images. The products obtained from the STScI 
archive are hence fully calibrated and distortion-corrected images, in units of counts per second.

\begin{figure*}
\begin{minipage}{175mm}
\begin{center}
\includegraphics[width=86mm]{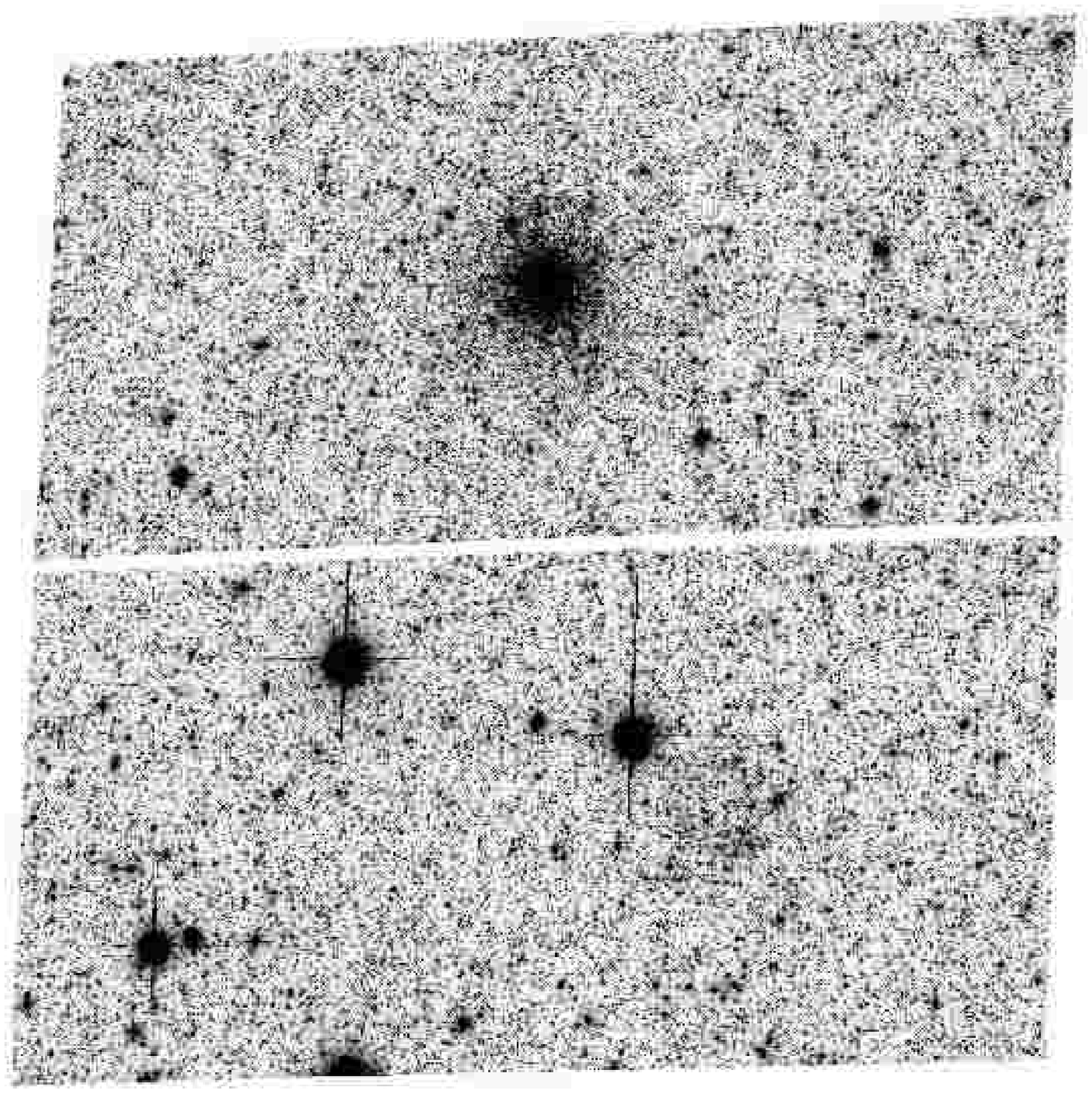}
\hspace{0mm}
\includegraphics[width=86mm]{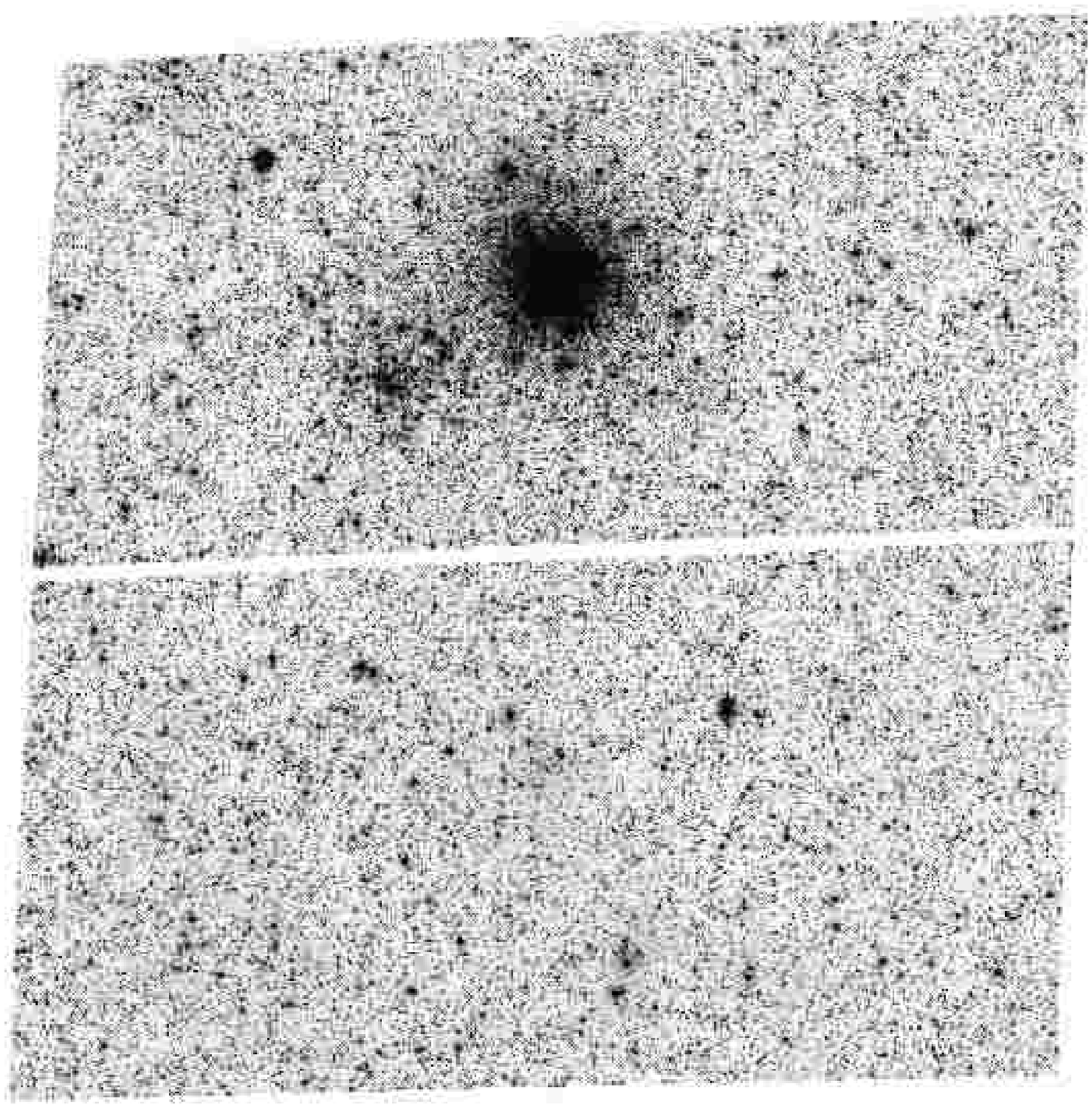} \\
\vspace{1mm}
\includegraphics[width=86mm]{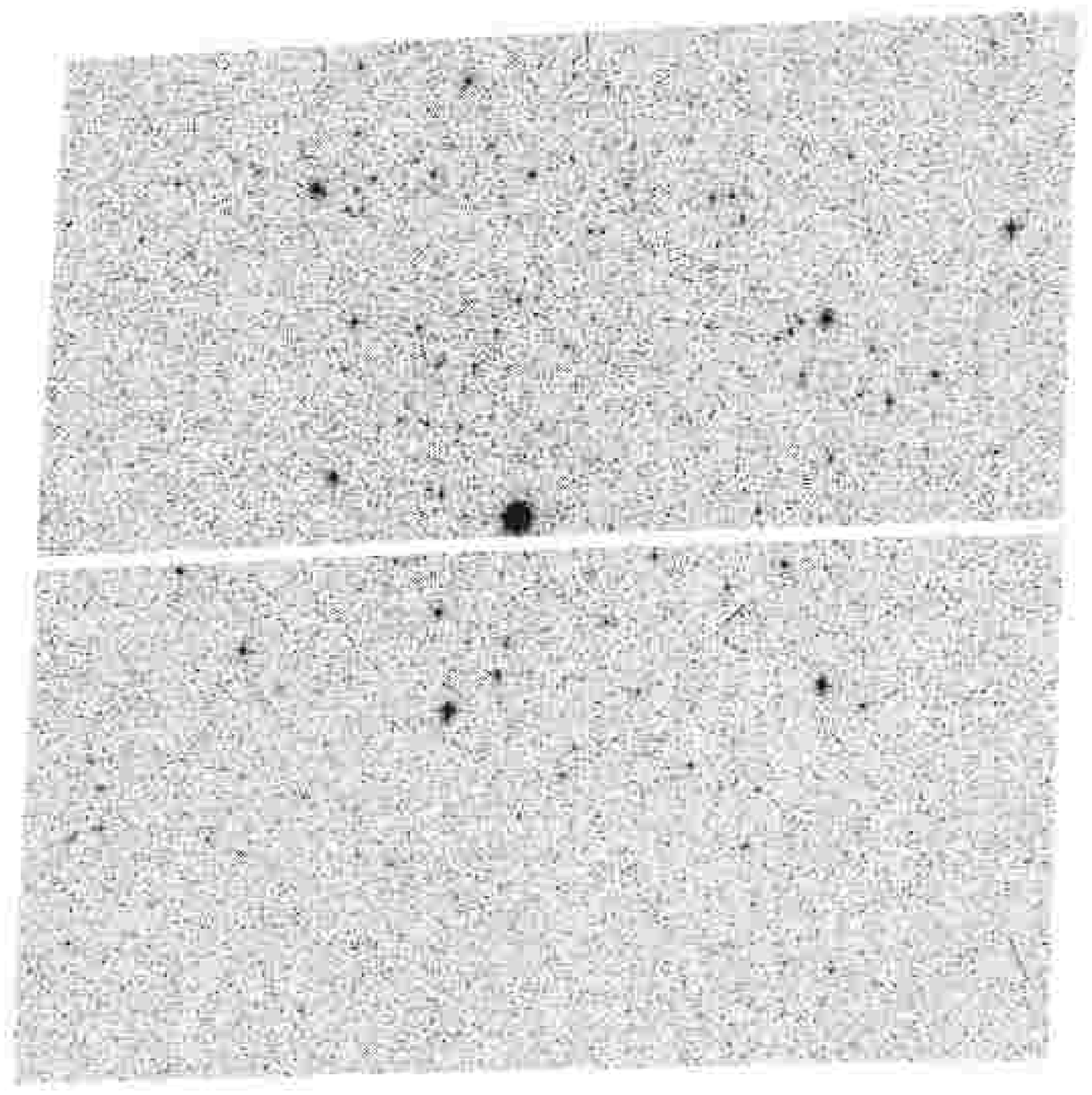}
\caption{Distortion-corrected F555W-band ACS/WFC images of NGC 1928 (upper left), NGC 1939 (upper right), and Reticulum (lower). Note the presence of a second cluster (NGC 1938) in the NGC 1939 frame -- this object lies to the lower left of the main cluster.}
\label{f:images}
\end{center}
\end{minipage}
\end{figure*}

The F555W images of NGC 1928, 1939, and Reticulum may be seen in Figure \ref{f:images}. Both
NGC 1928 and NGC 1939 are extremely compact clusters set against heavily populated background
fields, while Reticulum is at the other end of the scale -- extremely diffuse with almost no
background population evident. A second cluster is visible in the image of NGC 1939, slightly
to the lower left of the main cluster. This is another LMC member, NGC 1938 (see Section
\ref{ss:1938} and Figure \ref{f:1938}).

For each cluster, photometry was performed on the two images individually, using the {\sc daophot}
software in {\sc iraf}. The detailed procedure was as follows. First the {\sc daofind} task was 
used with a detection threshold of $4\sigma$ above background to locate all the brightness peaks
in each image. The two output lists were matched against each other to find objects falling at 
identical positions in the two frames. Objects detected in the first image but with no matching 
counterpart in the second image (and vice versa) were discarded. 

Performing this cross-matching was not as simple a task as it might at first appear. Because 
of the significant geometric distortion present in the ACS/WFC observations, a $2\times 2$ pixel 
offset in the telescope position (i.e., in the object's position on the detector) between the two 
exposures does not correspond to a $2\times 2$ pixel offset between the two distortion-corrected 
images. In fact, the value of the offset is position-dependent in the corrected images. There are 
at least two possible ways to overcome this. The simplest is to use the header information in each 
of the two images to overlay a common coordinate system (e.g., J2000.0 $(\alpha\,,\,\delta)$ 
coodinates) and match between the lists via a transformation from pixel coordinates to the common 
coordinate system. However, this procedure is completely dependent on the accuracy of the header 
information in each of the two datasets and the consistency between them, and we possessed no 
independent means of verifying this for all the observations. Instead, we preferred a more 
generally applicable and physically grounded procedure. We applied the distortion models used 
in {\em PyDrizzle} to transform the observed F555W positions from the distortion-corrected frame 
to detector positions (i.e., in the distorted frame). The new coordinates were then subjected to 
the $2\times 2$ pixel offset and transformed into the F814W distortion-corrected frame. This 
allowed a direct match against the list of object positions measured by {\sc daofind} from 
this frame. Given that the distortion models are accurate to a small fraction of a pixel, a 
match was defined as the transformed F555W position lying within a $0.8$ pixel radius of an F814W
position. Experimentation showed this limiting radius to be perfectly adequate across the full WFC 
field of view.

The model defining the relationship between detector coordinates and distortion-corrected 
coordinates takes the form of a polynomial transformation:
\begin{eqnarray}
x_c = \sum_{i=0}^{k} \sum_{j=0}^{i} a_{i,j} (x-x_r)^{j} (y-y_r)^{i-j} \\
y_c = \sum_{i=0}^{k} \sum_{j=0}^{i} b_{i,j} (x-x_r)^{j} (y-y_r)^{i-j}
\end{eqnarray}
where $(x\,,\,y)$ are the detector coordinates (in pixels), $(x_c\,,\,y_c)$ are the corrected 
coordinates (in arcseconds), $(x_r\,,\,y_r)$ is the position of a reference pixel, and $k$ is the 
order of the polynomial. The inverse transformation takes a similar form. Additional corrections 
are required to provide a grid common to the two chips which make up the WFC. Full details of the
model, including its derivation and application, are  provided by Mack et al. 
\shortcite{acsdatabook}. The latest solution may be downloaded from the STScI web site in the form 
of a matrix of the polynomial coefficients ($a_{i,j}$, etc) along with the required offsets, 
reference pixel values and plate scales. The version used in the present work was named 
nar11046j\_idc.

With the cross-matching complete, lists of detected objects were provided to the {\sc phot} task.
This was used to perform aperture photometry on each object, using apertures of radius $r = 3$ 
pixels. We also attempted to use {\sc daophot} routines to perform PSF-fitting photometry; however,
the quality and internal consistency of the measurements we obtained was not as good as for the
aperture photometry. The reasons for this likely have to do with both the nature of the images
and of the objects imaged. The PSF is apparently significantly variable across the WFC field of view
(possibly due to small imperfections in the distortion correction); while the observations of 
NGC 1928 and 1939 are extremely crowded (we measured more than $10^5$ detections in each field),
meaning that it is difficult to find suitable stars to construct model PSFs across the entire
field of view. The $3$ pixel aperture photometry radius is large enough to be relatively insensitive
to PSF variations and small enough to be usable given the crowding, so is an acceptable compromise.

The resultant photometry has been calculated in the STmag system, defined as 
$m = -2.5 \log_{10} f_\lambda - 21.1$, where $f_\lambda$ is the flux density per unit wavelength,
and the zero-point is set so that Vega has magnitude $0$ in the Johnson $V$ passband.
The constants required to convert the measured photometry from counts to $f_\lambda$ were
selected from the ACS zero-points web page, and correspond to applying the formula
$m = -2.5 \log_{10} ({\rm counts\,\,s}^{-1}) + ZP$, where $ZP = 25.672$ for the F555W filter and 
$ZP = 26.776$ for the F814W filter.

Like all previous {\em HST} CCD instruments, the ACS/WFC chips are suffering from degradation
of their charge transfer efficiency (CTE) due to radiation damage. Since ACS is a relatively
new instrument, this degradation is not yet large. Nonetheless, photometric measurements
already require correction to account for lost flux from imperfect CTE. A calibration of the losses 
due to parallel ($y$-direction) CTE effects for ACS/WFC has been provided by Riess \shortcite{cte}, 
who parametrizes the necessary correction by:
\begin{equation}
\Delta_Y = 10^A \times s^B \times f^C \times \frac{Y}{2048} \times \frac{(MJD - 52333)}{(52714 - 52333)}\,\,\,{\rm mag}
\label{e:cte}
\end{equation}
where the object's sky ($s$) and flux ($f$) values are in counts, $Y$ represents the number of 
parallel transfers (so if the object has position $(x\,,\,y)$, then $Y=y$ for $1 \le y \le 2048$ 
and $Y=4096-y$ for $y > 2048$), and MJD is the Modified Julian Date of the observation being 
corrected. The Riess calibration provides exponents $A = 0.45 \pm 0.10$, $B = -0.11 \pm 0.03$ and 
$C = -0.65 \pm 0.04$ for aperture photometry measurements with radius $r=3$ pixels. At present 
there appear to be no additional corrections required for serial ($x$-direction) transfer.

With the CTE corrections calculated, the final step was to determine aperture corrections for each 
of the measured images. In principle it is desirable to correct to an infinite aperture; however,
this was not possible here since no standard stars were observed in our images. Alternatively,
it is common to correct to a set aperture, after which a transformation which includes correction
to an infinite aperture is often applied to the measurements to place them on a standard magnitude 
scale (e.g., to move from F555W flight magnitudes to standard Johnson $V$ magnitudes). At the time 
of writing however, no such transformations are yet available for the ACS/WFC filter system. 
Nonetheless, it has become common practise for WFPC2 measurements to correct to an aperture of 
radius $0.5\arcsec$ as prescribed by Holtzman et al. \shortcite{holtzman}. We therefore calculated
a correction to this aperture ($10$ ACS/WFC pixels) for the present photometry. It is important
to note that the aperture correction does not affect the relative photometry for a given cluster
since it is applied uniformly across each set of measurements. It is in general important only for 
placing the photometry on an absolute scale. As an indication of the systematic absolute error,
the energy encirclement plots provided by Mack et al. \shortcite{acsdatabook} show that just under
$95$ per cent of the flux of a point source is contained within $0.5\arcsec$ radius. Once suitable
transformations for ACS/WFC photometry have been published, it will be a simple procedure to
re-calculate the corrections to the required aperture and convert to a standard magnitude system.

Aperture stars were selected according to strict criteria: they must be bright stars but 
significantly below saturation; they must not have unusual shape characteristics (see Section
\ref{s:colourmag}); they must have no neighbouring stars, bad pixels, or image edges within a 
radius of $1\arcsec$; and they must not lie in an area of unusually high background (e.g., near the 
centre of a cluster). For NGC 1928 and 1939 these criteria defined a set of several hundred stars
per image, while for Reticulum the set numbered approximately $100$ stars. The photometry
procedure described above was repeated on these star lists using apertures of radius
$10$ pixels, and the mean aperture correction for each image calculated using a $3\sigma$ clipping
algorithm. Standard errors in the calculated corrections were typically $\sim 0.01$ mag.

\section{Colour-Magnitude Diagrams}
\label{s:colourmag}
\begin{figure*}
\begin{minipage}{175mm}
\begin{center}
\includegraphics[width=87mm]{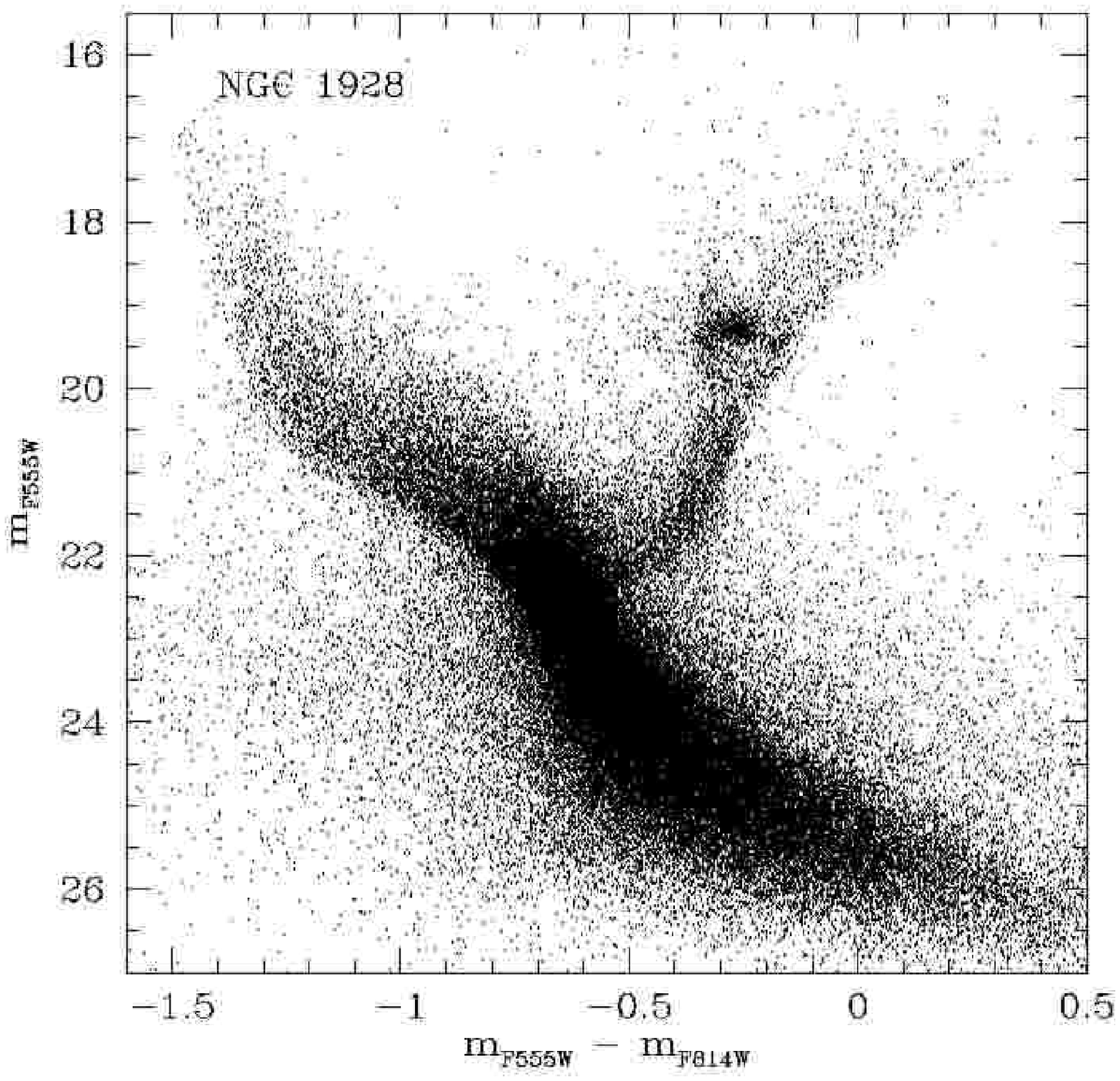}
\hspace{-2mm}
\includegraphics[width=87mm]{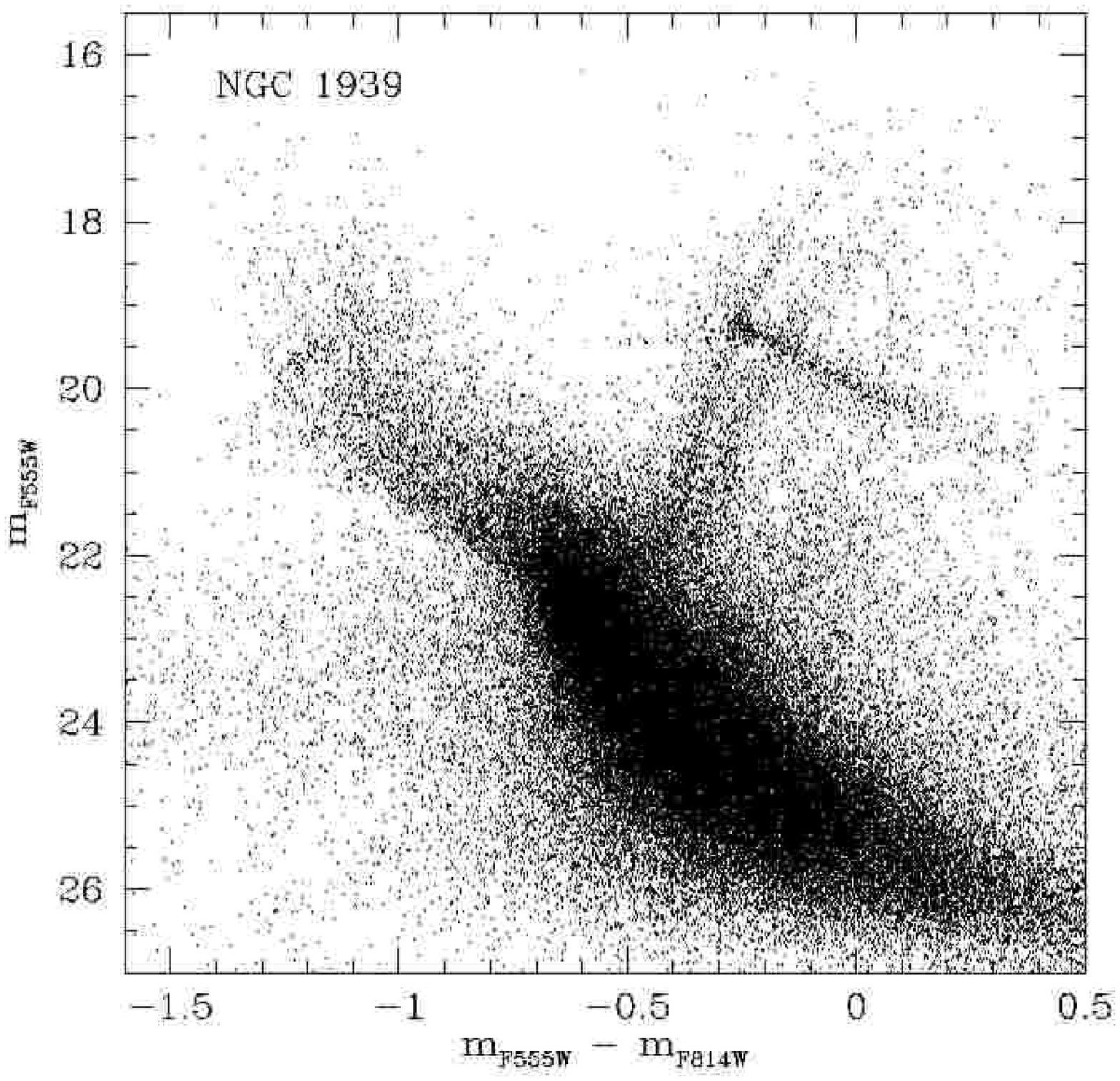} \\
\vspace{1mm}
\includegraphics[width=87mm]{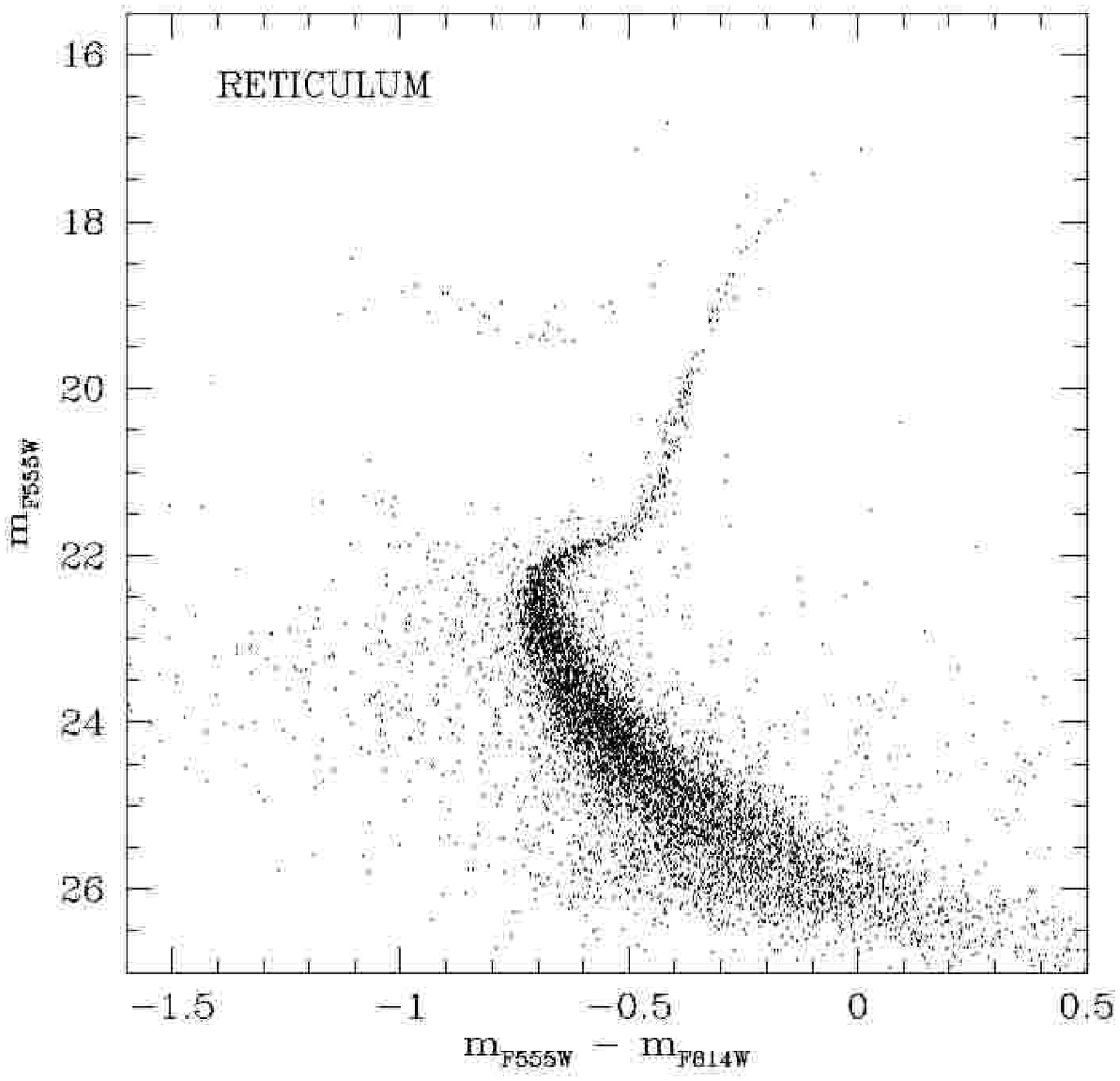}
\caption{Colour-magnitude diagrams for all detections in each of the three fields. Measurements are plotted in the STmag magnitude system (see text). The CMD for NGC 1928 contains $104\,438$ detections, while that for NGC 1939 contains $94\,546$ detections, and the CMD for Reticulum $7\,604$ detections.}
\label{f:cmdall}
\end{center}
\end{minipage}
\end{figure*}

Colour-magnitude diagrams for the three clusters are presented in Fig. \ref{f:cmdall}. These
diagrams show all matched detections -- there has been no selection of objects according to
shape characteristics (see below). It is clear that the diagrams for NGC 1928 and 1939 are 
dominated by field stars, as expected. NGC 1928 lies against a somewhat denser field than 
NGC 1939. Without some form of statistical subtraction, it is impossible to determine which 
parts of these two CMDs belong to the clusters. The NGC 1939 field suffers in addition from severe 
differential reddening, as is evident from the smearing of the red clump. In contrast, the CMD for
Reticulum is well defined and contains little or no field contamination. The narrowness of the 
sequences visible in all three CMDs (e.g., the red-giant branches) clearly demonstrate the high 
internal accuracy of the photometry and the validity of the reduction procedure described in the 
previous Section.

\subsection{Field Star Subtraction}
\label{ss:fieldsub}
While the CMD for Reticulum clearly possesses little or no field star contamination, statistical
subtraction of the field population was absolutely necessary before any study of NGC 1928 and 
NGC 1939 could be made. For such severely contaminated clusters, statistical subtraction is not a 
trivial matter. We developed two different subtraction methods and combined the results of each.

The first stage was to remove objects with unusual or non-stellar shape characteristics. Photometry 
for such objects is likely to be compromised (e.g., by a cosmic ray strike, bad pixel, or the blend 
of two or more very close stars), or the objects are likely to be background galaxies. Either way,
it is desirable to remove them from the CMDs. The output of the initial detection program,
{\sc daofind}, provided three shape characteristics per detection: a sharpness parameter and two
roundness parameters (called {\em s-round} and {\em g-round}). The sharpness is a measurement of
how high the peak of a detection is relative to a best-fitting Gaussian, while the two roundness
parameters are measurements of how circular the object image is. Clean stellar detections should 
have sharpness $\sim 0.75$ and round images (roundness parameters $\sim 0$). For each photometry
list we produced histograms of these three parameters in order to determine suitable clipping limits.
In general we removed objects which did not have $0.6 < {\rm sharpness} < 0.9$ and 
$-0.35 < {\rm roundness} < 0.35$ through both filters. This reduced the photometry lists for
NGC 1928 and NGC 1939 by $\sim 35$ per cent, and that for Reticulum by $\sim 50$ per cent 
(although for this cluster the removed objects were almost exclusively near the faint limit of 
detection).

The Reticulum photometry required no further subtraction. For NGC 1928 and 1939, each photometry
list was split into three groups according to radius from the centre of the cluster\footnote{We 
determined the two cluster centres by eye -- a procedure accurate enough for present purposes given
the very compact nature of both objects.}. The 
aim was to select two radii ($r_1$ and $r_2$) for each cluster, so that the detections within
$r_1$ defined as clean a cluster sample as possible, while those outside $r_2$ defined a clean
field sample. The sample in between $r_1$ and $r_2$ consisted of a mixed field/cluster sample.
A small amount of experimentation allowed appropriate values for these radii to be chosen.
To define $r_1$, photometry lists were assembled for all stars inside $r = 5$--$25\arcsec$ at 
$1\arcsec$ intervals. CMDs were constructed for each of these lists, and $r_1$ selected to
produce a well defined CMD with as little field contamination as possible. For NGC 1928, 
$r_1 = 12\arcsec$, while for NGC 1939 $r_1 = 15\arcsec$. These results are consistent with the
fact that NGC 1928 is somewhat more compact that NGC 1939 (see Fig.\ref{f:images}), as well as
the fact that NGC 1928 is set against a somewhat denser field than NGC 1939 (see Fig. 
\ref{f:cmdall}) -- both of which suggest that $r_1$ should be smaller for NGC 1928 than NGC 1939.
We note that for these measurements and all subsequent calculations, all stars within 
$12\arcsec$ of the centre of NGC 1938 on the NGC 1939 frames were excluded since a large fraction
do not belong either to NGC 1939 or its background field. The characteristics of this cluster are 
described briefly in Section \ref{ss:1938}.

Selecting radius $r_2$ was a matter of estimating the tidal radius ($r_t$) for each cluster. 
Calculating the surface density of stars in radial bins of width $5\arcsec$ from $50$--$140\arcsec$ 
showed the density to be approximately constant beyond $\sim 90\arcsec$ for NGC 1928 and 
$\sim 100\arcsec$ for 
NGC 1939. These measurements are consistent with previous measurements for other compact LMC bar 
clusters. For example Olsen et al. \shortcite{olsen} constructed King models for NGC 1754, 1835, 
2005, and 2019, and found $r_t \sim 115$, $85$, $105$, and $145\arcsec$ respectively. Similarly, 
Mateo \shortcite{mateo} measured $r_t \sim 50$, $145$, and $160\arcsec$ for NGC 1754, 1786, and 
1835, respectively. Elson \shortcite{elson:85} presented a high quality surface brightness profile
and King model for NGC 1835 (the most massive of the old LMC bar clusters, see Mackey \& Gilmore
\shortcite{sbp1}), and measured $r_t \sim 190\arcsec$ formally from the model. However, her
profile (her Figure 4a) shows the surface brightness to be greater than $10$ mag below the central
value even by $r \sim 100\arcsec$. Hence, we set $r_2 = 100\arcsec$ for NGC 1928 and 
$r_2 = 110\arcsec$ for NGC 1939. While these radii are likely smaller than the true tidal radii for
these clusters, they were perfectly adequate for present purposes -- allowing field samples of
$22\,000$ and $17\,000$ stars, respectively. The field star density was found to be $1.48$ 
arcsec$^{-2}$ for NGC 1928, and $1.29$ arcsec$^{-2}$ for NGC 1939.

We subjected the photometry samples between $r_1$ and $r_2$ to two statistical subtraction 
procedures. The first involved using the central (cluster) CMD to subtract a matching CMD
from the sample, leaving the field stars. The necessary number of subtractions was calculated by 
using the measured field star density to estimate the expected total number of field stars in the
region between $r_1$ and $r_2$. This in turn defined the expected number of cluster stars in the
region -- that is, the required number of subtractions. We did not account for detection completeness
in these estimates. Since regions closer in to the cluster have higher stellar densities, the
detection completeness decreases with decreasing radius. The completeness is also a function of 
stellar magnitude and colour. Because the density of detected field stars in the region between 
$r_1$ and $r_2$ is, overall, less than in the outer region, assuming full completeness in this area
leads to an over-estimate of the number of field stars in the intermediate region and hence an
under-estimate of the cluster population. However, this was not a significant problem for the 
present work, primarily because the area suffering from the greatest incompleteness is within
$r_1$ for both clusters. For radii greater than $r_1$ we estimate the completeness to be greater
than $90$ per cent for all areas of interest on the CMD (i.e., excluding the lower main sequences,
where photometric errors are, in any case, large). In addition, our primary aim was not to obtain
a subtraction of {\em all} possible cluster stars. Rather, we wanted to observe clean CMDs for
NGC 1928 and 1939 for the first time, for the purposes of photometric study. Had we been interested
in total number counts (as we would be in the construction of a brightness profile, for example)
full artificial star tests would have been carried out (see Mackey \& Gilmore, in prep.).

The subtraction process was as follows. First, a random star from the central region CMD was 
selected. On the intermediate region CMD, all non-subtracted stars within a $3\sigma$ error
ellipse from this star were located, and one randomly subtracted. If there were no stars within
the ellipse, the nearest neighbour was subtracted, providing it did not lie unreasonably distant
from the point (i.e., not more than $\sim 6\sigma$). This process was repeated the required number 
of times to obtain a realization of the subtracted cluster CMD. In any such realization, multiple 
selections of a central region star were allowed, but a star could not be subtracted from the 
intermediate region more than once.

For both NGC 1928 and 1939, one hundred CMD realizations were calculated. With these complete,
each star in the intermediate region was checked to find how many times it had been subtracted.
Stars with a large number of subtractions to their name were most likely to be cluster members.
For both clusters, a histogram of this statistic was constructed, allowing a suitable cut-off
to be estimated. A small amount of experimentation showed that selecting all stars with more
than $75$ subtractions provided clean, well-defined CMDs.

The second subtraction method was very similar, but involved using the outer (field) CMD to 
subtract a matching CMD from the intermediate sample, leaving the cluster stars. The required
number of subtractions was again determined using the measured field star density in the outer
region. As before, detection completeness was not accounted for, meaning the estimated number
of field stars (and hence subtractions) was over-estimated. The subtraction process was identical,
except this time random stars from the outer region CMD were selected. Once again, one hundred
realizations of each cluster's CMD were obtained and the stars appearing the most times in these
CMDs selected as the most likely cluster members. For this method, selecting all stars with
more than $50$ subtractions provided good CMDs. Because of the differential reddening present
in the outer field regions of the NGC 1939 frame, this subtraction method was not as effective
for this cluster. Nonetheless, adequate results were obtained.

\subsection{Final Cluster CMDs}
\label{ss:cmdgood}

\begin{figure*}
\begin{minipage}{175mm}
\begin{center}
\includegraphics[width=87mm]{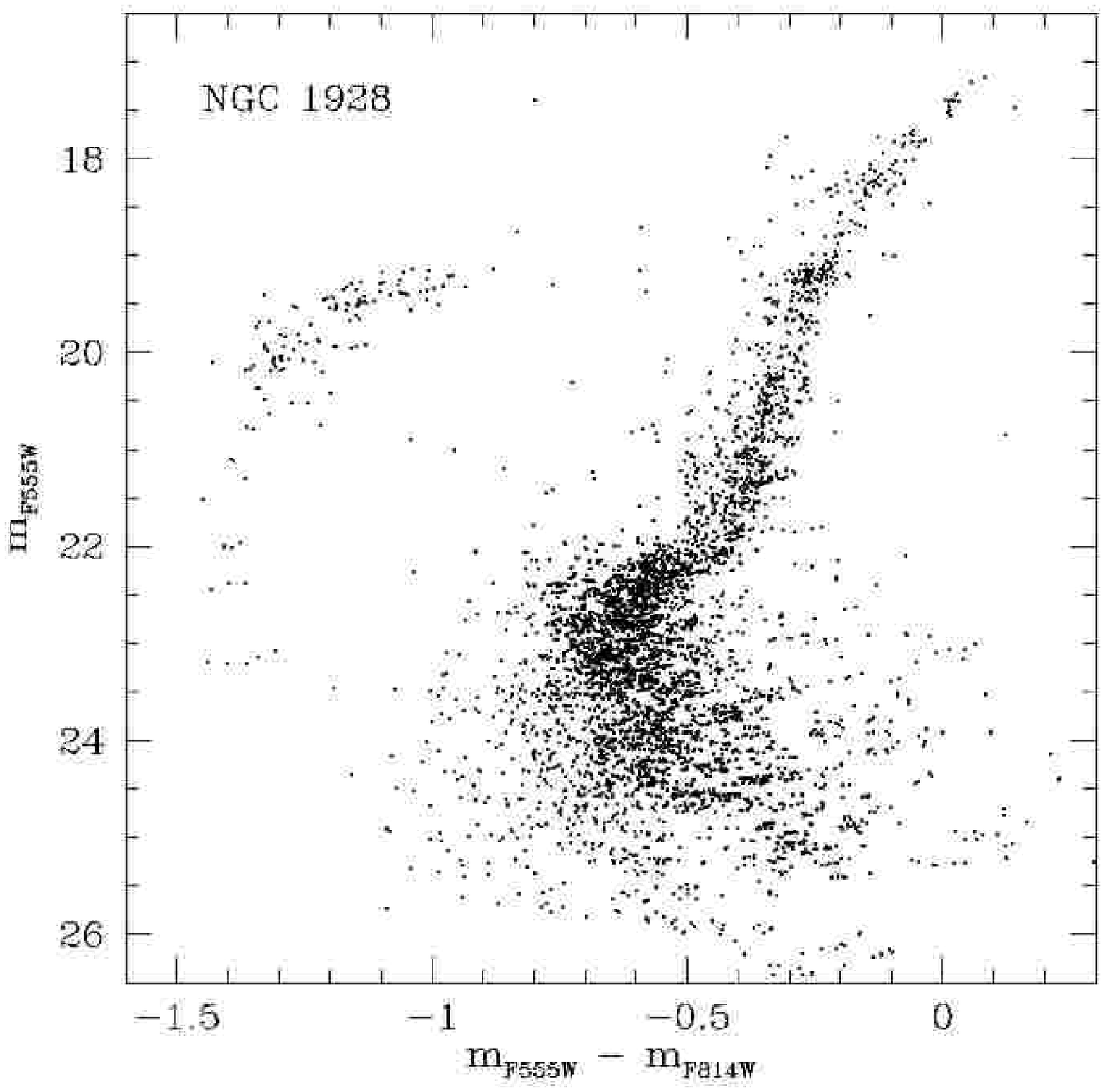}
\hspace{-2mm}
\includegraphics[width=87mm]{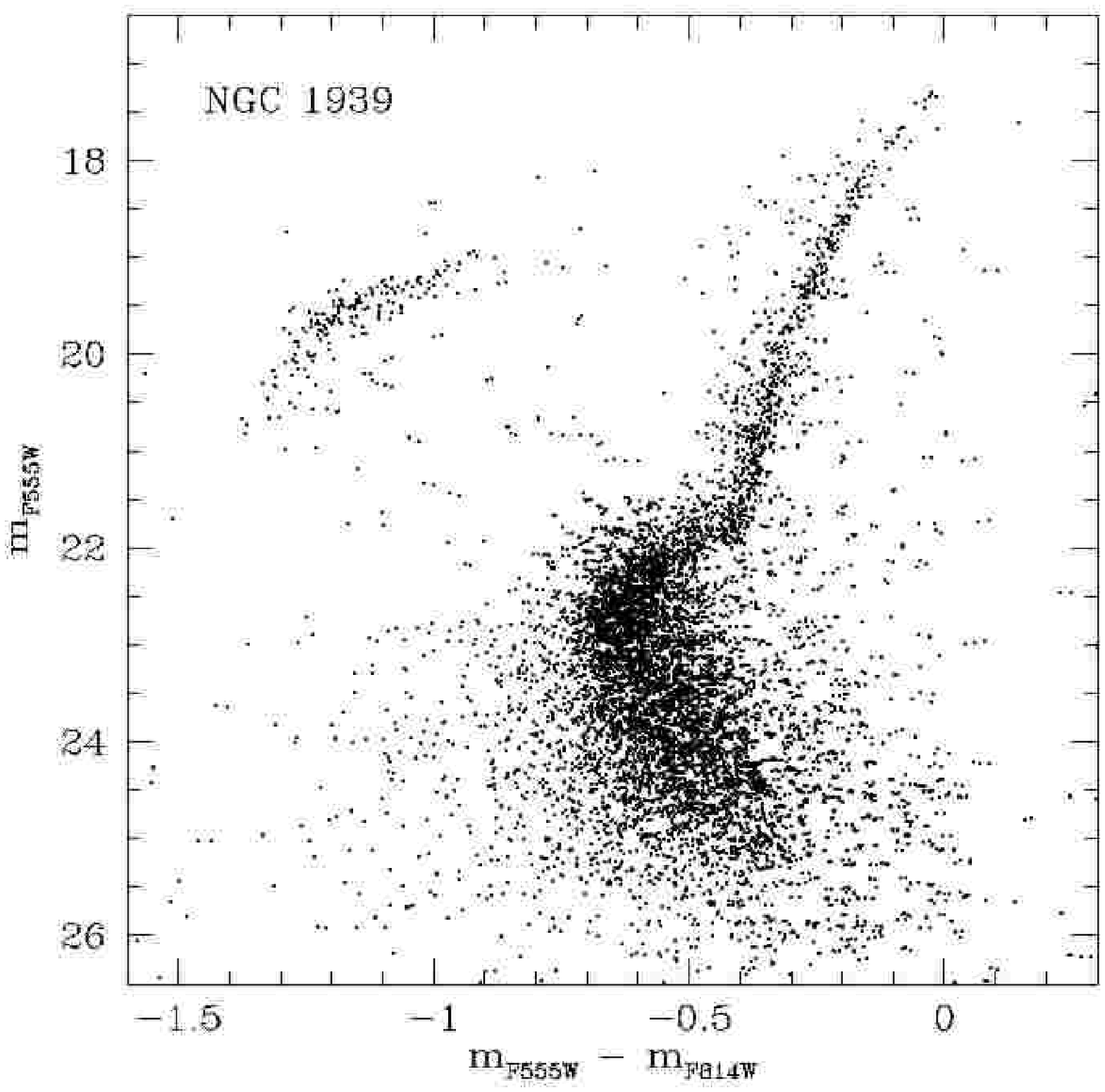} \\
\vspace{1mm}
\includegraphics[width=87mm]{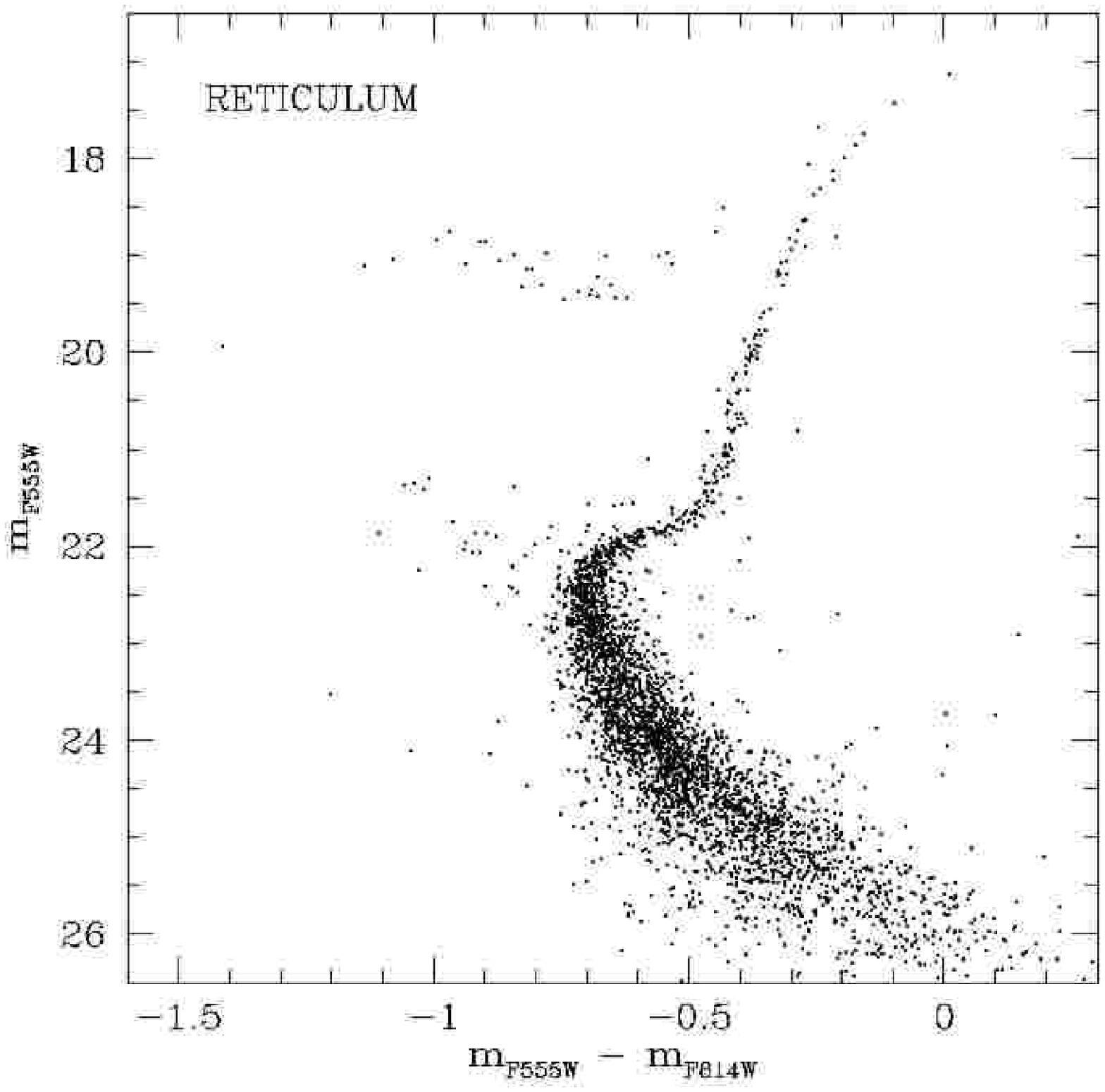}
\caption{Cleaned, field-subtracted colour-magnitude diagrams for the three clusters. For both NGC 1928 and 1939, the plotted points are the combined results of the two field subtraction algorithms, along with the central sample of stars.}
\label{f:cmdsub}
\end{center}
\end{minipage}
\end{figure*}

The final cleaned, field-subtracted CMDs for NGC 1928, 1939, and Reticulum appear in Fig. 
\ref{f:cmdsub}. In this Figure, the CMDs for NGC 1928 and 1939 represent the combined results
of the two subtraction processes. Stars within $r_1$ have also been plotted for both clusters,
since these are very predominantly cluster members. To the best of our knowledge these are the
first published CMDs for NGC 1928 and 1939. It can clearly be seen that these two clusters are 
very old, thus confirming the results obtained by Dutra et al. \shortcite{dutra} using integrated 
spectroscopy. Both clusters appear to possess well populated horizontal branches, consisting
almost entirely of blue stars. In this respect they strongly resemble the old LMC clusters
Hodge 11 \cite{walker:h11,mighell,johnson:99} and NGC 2005 \cite{olsen}. In addition, NGC 1928 
apparently possesses an extended blue HB, falling to $V\sim 23$. The effects of differential 
reddening on the field subtraction for NGC 1939 can be seen in its lower main sequence, which 
exhibits a sharp cut-off to the red. The remainder of the CMD has not been significantly affected 
by this problem.

Reticulum is also seen to be an old cluster, with a horizontal branch primarily consisting of stars 
within and to the red of the instability strip. A significant population of blue stragglers also 
appears to be present. This CMD confirms the earlier results of Walker \shortcite{walker:ret}; 
Johnson et al. \shortcite{johnson:02}; Marconi et al. \shortcite{marconi}; and Monelli et al. 
\shortcite{monelli}. 

\begin{figure}
\includegraphics[width=0.5\textwidth]{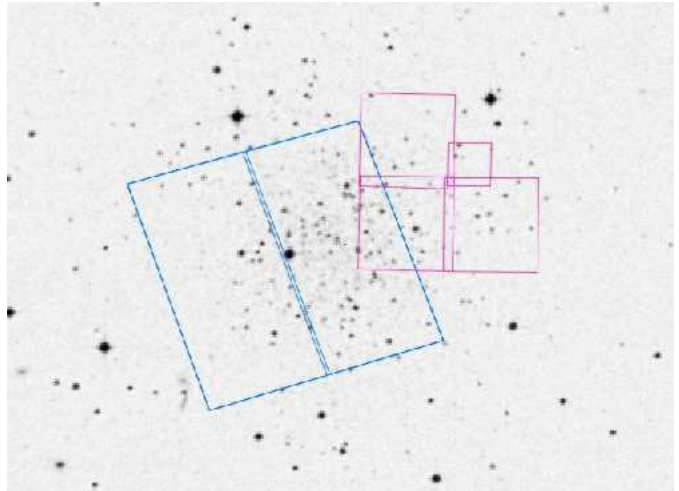}
\caption{Position and orientation of the present ACS observations of Reticulum and the archival WFPC2 observations, superimposed on a DSS image. North is toward the top and east to the left. For some idea of the scale, the ACS FOV is approximately $202\arcsec$ on a side. There is a small overlap between the two sets of observations, but the bulk of the WFPC2 field lies more than an arcminute to the north-west of the ACS frames.}
\label{f:ret}
\end{figure}

Ideally, we would like to use the final CMDs to provide photometric measurements of the cluster
reddenings and metallicities, and to place some constraints on their ages. However, such 
calculations generally require the photometry to be on a standard magnitude scale -- in this case, 
Johnson-$V$ and Cousins-$I$. At the time of writing, no transformations from the ACS/WFC STmag 
system to the Johnson-Cousins $VI$ system are yet available. Nonetheless, we were able to
determine an approximate transformation. One of the clusters from the present study -- Reticulum -- 
has previously been observed with {\em HST}/WFPC2 through the F555W and F814W filters (as part of
program 5897). Calibrations for these filters to the Johnson-Cousins $VI$ system {\em do} 
exist and are well established \cite{holtzman,dolphin}. The relevant archived data-groups are 
labelled u2xj0605b ($5 \times$ F555W frames -- two with exposure durations of $260$ s and three with
exposure durations of $1000$ s) and u2xj0608b ($6 \times$ F814W frames -- two with exposure 
durations of $260$ s and four with exposure durations of $1000$ s). Unfortunately, these 
observations did not image the cluster core, but rather are centred approximately $1.5\arcmin$ to 
the north west. Fig. \ref{f:ret} shows the positions of the two observation sets overlaid on a DSS 
image of Reticulum. While there is a small overlap, the number of common stars was not large enough 
to derive a point-by-point photometric transformation. Nonetheless, it was still possible to 
calculate a global transformation. 

\begin{figure}
\includegraphics[width=0.5\textwidth]{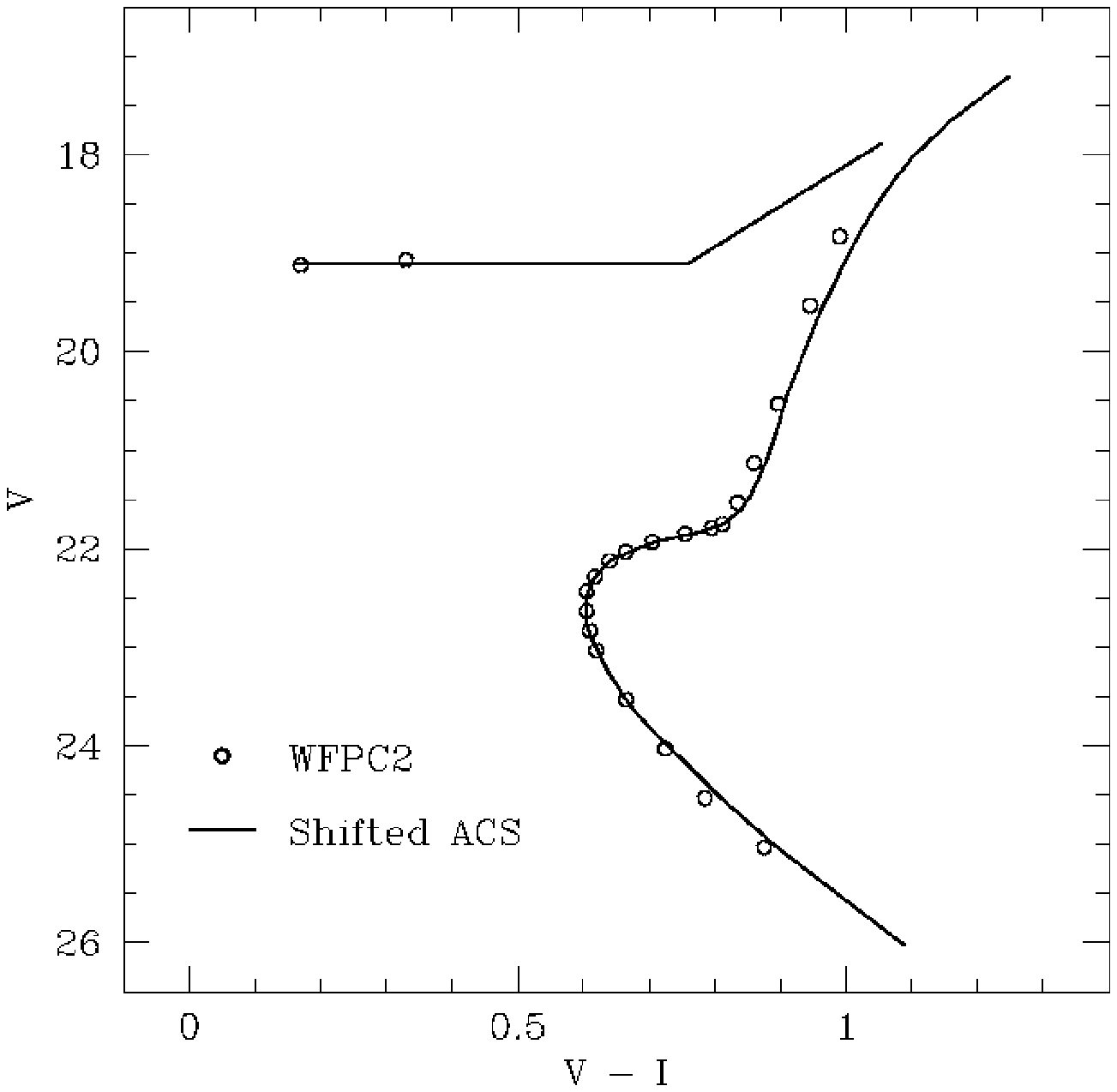}
\includegraphics[width=0.5\textwidth]{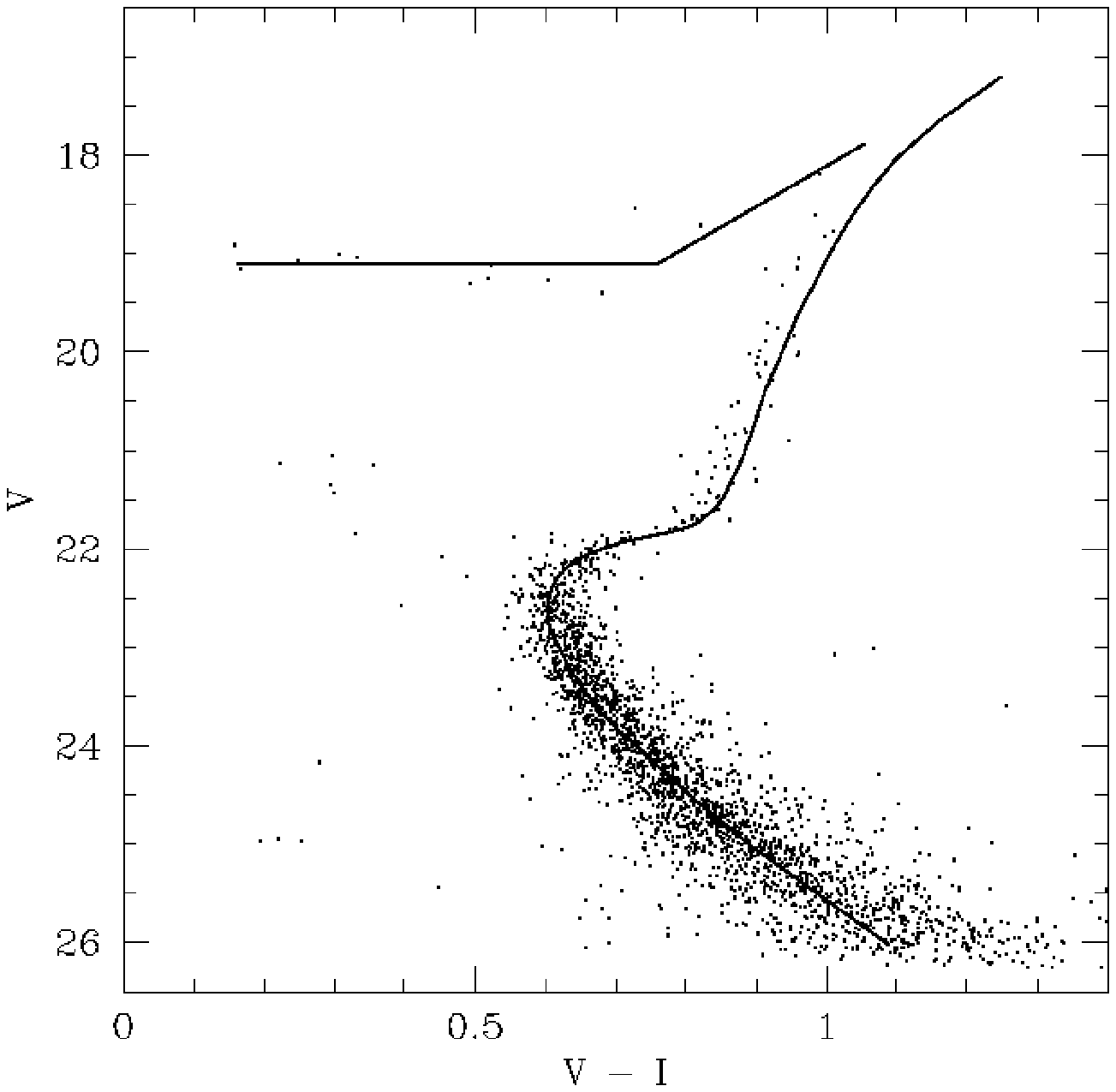}
\caption{The shifted Reticulum fiducial as determined from ACS/WFC photometry, compared with the Reticulum fiducial from the WFPC2 photometry (upper panel) and the WFPC2 photometry itself (lower panel). The ACS/WFC fiducial has been moved $0.03$ mag fainter in $V$ and $1.31$ mag redder in colour, as described in the text.}
\label{f:retmatch}
\end{figure}

First, photometric measurements were performed on the archival WFPC2 images using {\sc hstphot}
\cite{hstphot}. The measurement procedure is described fully by Mackey \shortcite{thesis},
and is identical to the procedure used by Mackey \& Gilmore \shortcite{rrlyr} to measure 
RR Lyrae stars in the globular clusters of the Fornax dwarf galaxy. The resultant CMD may be seen
in Fig. \ref{f:retmatch}. 

We next determined fiducials for both the CMD from the present study (in the ACS/WFC STmag system)
and the CMD from the archival WFPC2 data (in $V$, $V-I$). The main sequences are well populated
enough that the fiducials below the turn off could be calculated by forming magnitude bins and
finding the mode in colour for each. This would typically also work for the RGB, but on neither CMD
is this region particularly well populated (Reticulum is a very sparse cluster). Thus, the RGB
fiducials were determined by eye, as were the fiducials around the turn-off and sub-giant branch
(SGB). These latter could not be measured via a simple binning technique because they are neither
horizontal nor vertical. As can be seen in Figs. \ref{f:cmdsub} and \ref{f:retmatch}, the dispersion
in these regions of each CMD is small (particularly for the ACS measurements), so the by-eye
procedure should not introduce any large errors. This algorithm is very similar to many used in
previous studies (see e.g., Johnson et al. \shortcite{johnson:99}). The horizontal branch (HB)
levels were determined using the (very few) stars just to the blue of the instability strip.
Due to the lack of multi-epoch observations (especially for the ACS data), the RR Lyrae regions
possess a significant spread in magnitude due to the intrinsic stellar variability.

On each CMD, the colour of the main sequence turn-off (MSTO) was determined by fitting a 
second-order polynomial to the the data in this region, and finding the bluest point of the fit. 
The magnitude of a main-sequence (MS) reference point $V_{0.05}$ (the point on the MS which is
$0.05$ mag redder than the MSTO) was then determined by interpolating along the fiducial line.
These two values are traditionally used to register cluster CMDs for the purposes of differential
age comparison \cite{vbs} (see Section \ref{ss:ages}); however our purpose here was to determine
a linear shift between the ACS/WFC photometry, and the standard Johnson-Cousins scale. By moving
the ACS/WFC fiducial so that the colour of its MSTO matched that for the WFPC2 fiducial, and its
$V_{0.05}$ matched that for the WFPC2 fiducial, this transformation was calculated. We found that
$V - m_{{\rm F555W}} = 0.03$ and $(V-I) - (m_{{\rm F555W}}-m_{{\rm F814W}}) = 1.31$. The registered 
fiducials may be seen in the top panel of Fig. \ref{f:retmatch}, while the shifted ACS fiducial is 
plotted on the WFPC2 CMD in the lower panel of this Figure. The registration is very close across 
all parts of both fiducials, except in the RGBs, which are separated slightly. The upper portion of 
the WFPC2 RGB is somewhat uncertain due to a lack of stars; however the separation persists into
the lower RGB area, which is well defined for both sets of photometry. The HB levels, on the other
hand, match closely. Our derived transformation implies that $I - m_{{\rm F814W}} = -1.28$, which 
matches perfectly the result of Brown et al. \shortcite{brown}, who found exactly this relation for 
a $5000$ K stellar spectrum. 

While it is almost certain that the true transformation is not linear (the WFPC2 transformations
are parametrized by quadratic functions -- see Holtzman et al. \shortcite{holtzman}), it is clear 
from the good match between the two Reticulum fiducials that the present approximation is a good 
one, and certainly accurate enough for our purposes. It is not clear why the two RGBs show a small 
offset ($\sim 0.02$ mag). It is possible that this is a second order distortion (e.g., a stretch 
along the colour axis). If this were the case, we would expect a similar offset between the lower 
main sequences, which lie at approximately the same colour as the RGBs. There is indeed a hint of 
such an offset between the two sequences on the lower MS, although it is noted that the photometric 
errors are large at these faint magnitudes. We would also expect the discrepancy to become larger
with redder colours, and again, while there is a hint of this on the RGBs, the upper two points on
the WFPC2 fiducial are quite uncertain. Nonetheless, the possibility of second-order distortions 
(of the order of $0.02$ mag) must be considered in any calculations involving the ACS/WFC CMDs, 
such as those in the next Section, until the full ACS/WFC calibration is complete and comprehensive 
transformation equations are published.

\section{Cluster Properties}
\label{s:clusterprop}

\subsection{Abundances and Reddenings}
\label{ss:metred}
\begin{figure*}
\begin{minipage}{175mm}
\begin{center}
\includegraphics[width=45mm]{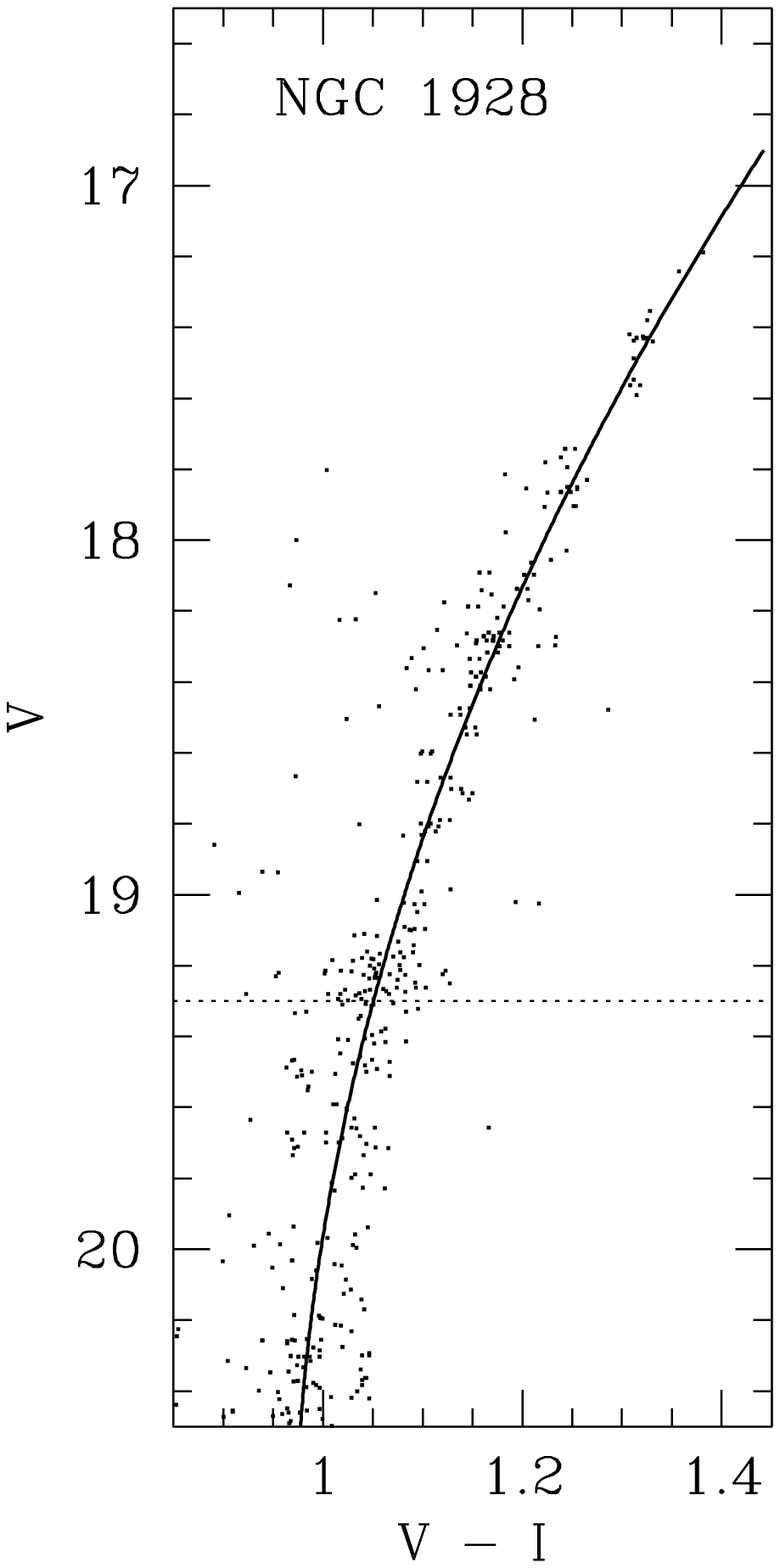}
\hspace{2mm}
\includegraphics[width=45mm]{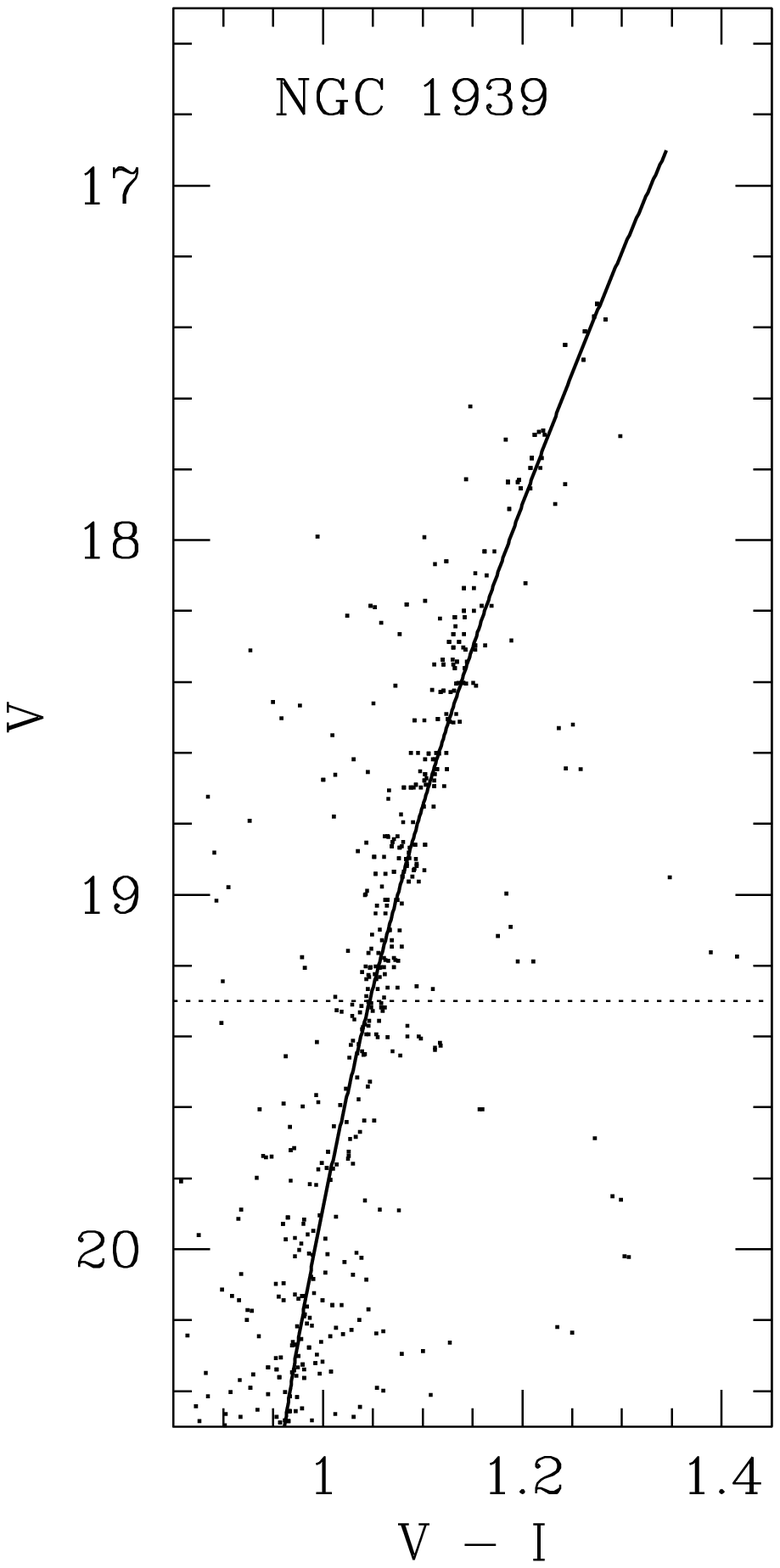}
\hspace{2mm}
\includegraphics[width=45mm]{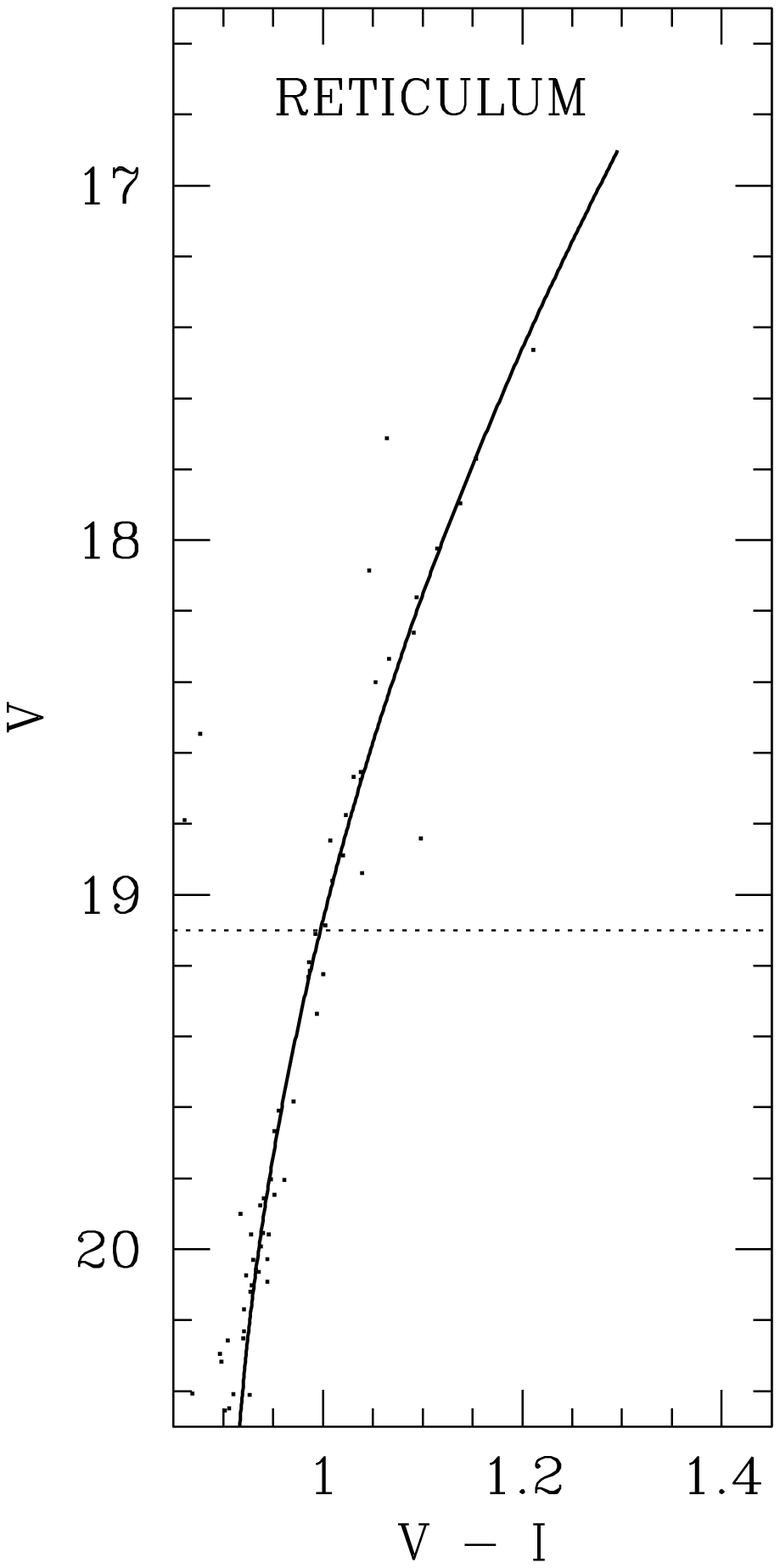}
\caption{Quadratic fits to the three cluster RGBs, used for the reddening and metallicity determinations. The dotted lines indicate the measured HB levels.}
\label{f:polyfit}
\end{center}
\end{minipage}
\end{figure*}

We employed the technique of Sarajedini \shortcite{s94} to calculate photometric estimates for the
reddening and metallicity of each of the three clusters. In this technique, the height of the RGB
above the HB at intrinsic (dereddened) colour $(V-I)_0 = 1.2$ is used to determine the metallicity
(this parameter is labelled $\Delta V_{1.2}$), while the reddening is obtained from the colour of 
the RGB at the level of the HB, $(V-I)_g$. The input parameters are hence the
$V$ magnitude of the HB and some parametrization of the shape of the RGB -- we chose to use the 
common procedure of fitting a second order polynomial to the RGB points above the HB. This took
the form $(V-I) = a_0 + a_1V + a_2V^2$. We fit this relation iteratively, discarding outlying
points on each iteration. The results may be seen in Table \ref{t:metred}, and graphically in 
Fig. \ref{f:polyfit}.

Measuring accurately the level of the HB ($V_{{\rm HB}}$) was not a trivial procedure, especially
for NGC 1928 and NGC 1939, which have predominantly blue HB morphologies, and unknown reddenings. 
Ideally in such cases, we would like to use the level of the reddest HB stars on the blue side of 
the instability strip. Using the RR Lyrae stars in four globular clusters in the Fornax dwarf 
spheroidal galaxy, Mackey \& Gilmore \shortcite{rrlyr} found the blue edge of the instability strip 
to lie at an intrinsic colour of $(V-I)_{{\rm BE}} = 0.28 \pm 0.02$. We employed an iterative 
technique to 
determine $V_{{\rm HB}}$ in conjunction with $E(V-I)$ and $[$Fe$/$H$]$ for each cluster. We first
determined estimates for $V_{{\rm HB}}$ using the reddest end of the populated blue HB region
for NGC 1928 and 1939, and the bluest HB stars for Reticulum. For a given cluster, we then solved
for $E(V-I)$ and $[$Fe$/$H$]$ following Sarajedini \shortcite{s94}, and used the value of 
$E(V-I)$ so determined to locate the HB stars just to the blue of the edge of the instability
strip at $(V-I)_0 = 0.28$. These stars defined a new value for $V_{{\rm HB}}$, and the process
was iterated until convergence. We estimated the (random) errors in these values again using 
the proscription of Sarajedini \shortcite{s94}. The accuracy with which we could measure 
$V_{{\rm HB}}$ was generally $\pm 0.05$ mag, while that for $(V-I)_g$ was greater -- because of
the narrowness of the RGB sequences and the stability of the quadratic fits -- at $\pm 0.01$ mag.
Ten thousand new fits per cluster were calculated, each time using a value of $V_{{\rm HB}}$
chosen randomly from a distribution with $\sigma = 0.05$ about the genuine measurement of 
$V_{{\rm HB}}$, and a new value of $(V-I)_g$ selected randomly from a distribution with
$\sigma = 0.01$ about the genuine measurement of $(V-I)_g$. The standard deviations in the new
sets of $E(V-I)$ and $[$Fe$/$H$]$ defined the random errors in these quantities.

\begin{table*}
\begin{minipage}{155mm}
\caption{Results of the simultaneous determination of cluster reddenings and metallicities.}
\begin{tabular}{@{}lccccccccc}
\hline \hline
Cluster & \hspace{3mm} & $V_{{\rm HB}}$ & $a_0$ & $a_1$ & $a_2$ & \hspace{3mm} & $\Delta V_{1.2}$ & $[$Fe$/$H$]$ & $E(V-I)$ \\
\hline
NGC 1928 & & $19.30 \pm 0.05$ & $13.45040$ & $-1.18991$ & $0.02836$ & & $1.64 \pm 0.15$ & $-1.27 \pm 0.14$ & $0.08 \pm 0.02$ \\
NGC 1939 & & $19.30 \pm 0.05$ & $8.22729$ & $-0.65527$ & $0.01468$ & & $2.52 \pm 0.20$ & $-2.10 \pm 0.19$ & $0.16 \pm 0.03$ \\
Reticulum & & $19.10 \pm 0.05$ & $10.56360$ & $-0.91361$ & $0.02161$ & & $2.05 \pm 0.13$ & $-1.66 \pm 0.12$ & $0.07 \pm 0.02$ \\
\hline
\label{t:metred}
\end{tabular}
\end{minipage}
\end{table*}

The final reddening and metallicity values are recorded in Table \ref{t:metred} along with the 
estimated errors. NGC 1939 is a metal-poor cluster ($[$Fe$/$H$] = -2.10$), while Reticulum is somewhat
more metal-rich ($[$Fe$/$H$] = -1.66$). In contrast, NGC 1928 is significantly more metal-rich again
with $[$Fe$/$H$] = -1.27$ -- rendering it the most metal-rich of the known old LMC bar clusters.
Examination of the clean CMD for NGC 1928 supports this result, as there is a clear RGB luminosity
function bump at approximately $V_{{\rm HB}}$. Such a bump is characteristic of clusters with 
intermediate metal abundance (see e.g., Sarajedini \& Forrester \shortcite{sarajedini:95}). 

Our new results are all consistent with previous measurements and estimates, where these
are available. NGC 1928 and 1939 have been poorly studied, with each possessing only one previous
metallicity estimate. These are from Dutra et al. \shortcite{dutra} who compared their integrated
spectra of NGC 1928 and 1939 to those for three Galactic globular clusters to estimate that
$[$Fe$/$H$] \approx -1.2$ for NGC 1928, and $[$Fe$/$H$] \approx -2.0$ for NGC 1939. According
to Burstein \& Heiles \shortcite{dust}, the foreground reddening in the direction of both
is $E(B-V) = 0.09$, which corresponds to $E(V-I) \approx 0.12$ (see e.g., Mackey \& Gilmore 
\shortcite{fnx}). Reticulum has been more extensively studied. Walker \shortcite{walker:ret}
found that $[$Fe$/$H$] = -1.7 \pm 0.1$ by studying the RR Lyrae stars in the cluster, while 
Suntzeff et al. \shortcite{suntzeff} obtained a spectroscopic measurement of 
$[$Fe$/$H$] = -1.71 \pm 0.1$. Both of these measurements are in excellent agreement with our 
photometric determination that $[$Fe$/$H$] = -1.66 \pm 0.12$. On the reddening front, we measured
$E(V-I) = 0.07 \pm 0.02$. Walker \shortcite{walker:ret} found that $E(B-V) = 0.03 \pm 0.02$ 
(i.e., $E(V-I) = 0.04 \pm 0.03$) but notes that the colours of his RR Lyrae sample at minimum light 
could imply a slightly higher reddening: $E(B-V) = 0.05 \pm 0.02$ (i.e., $E(V-I) = 0.07 \pm 0.03$). 
Marconi et al. \shortcite{marconi} suggest that the reddening towards Reticulum could be twice
as large as that suggested in the literature (i.e., $E(V-I) \sim 0.08$). All of these estimates
are in good agreement with our new measurement.

As a final consistency check, if we adopt the calibration between intrinsic RR Lyrae brightness
and metallicity of Chaboyer \shortcite{chaboyer}: 
\begin{equation}
M_V({\rm RR}) = 0.23([{\rm Fe}/{\rm H}] + 1.6) + 0.56\,\,,
\end{equation}
we can calculate distance moduli ($\mu$) for the three clusters. We find that, with 
$A_V = 2.37E(V-I)$ (e.g., Mackey \& Gilmore \shortcite{fnx}), $\mu = 18.47 \pm 0.12$ for 
NGC 1928 and $\mu = 18.48 \pm 0.16$ for NGC 1939, where the errors represent only the
effect of random measurement errors from $V_{{\rm HB}}$, $E(V-I)$ and $[$Fe$/$H$]$. For Reticulum
we find $\mu = 18.39 \pm 0.12$. All these estimates are in good agreement with the canonical
distance to the LMC: $\mu_{{\rm LMC}} \approx 18.50$.

Given the uncertainties in the (approximate) photometric transformations detailed in Section
\ref{ss:cmdgood}, it is important to discuss briefly the potential effect of these on our 
metallicity and reddening measurements. The metallicity determination is obtained from a purely
differential process (Eq. 4 in Sarajedini \shortcite{s94} shows $[$Fe$/$H$]$ to be dependent only 
on $\Delta V_{1.2}$) and is thus affected only by any second or higher order distortion
over the range $\Delta V_{1.2}$ in $V$ and $1.2 - (V-I)_g$ in $V-I$. Since both of these,
especially the latter, are relatively small values, these distortions are unlikely to be large,
and we estimate the systematic error so introduced to be less than $0.05$ dex. Since the 
distortion apparently produces an RGB which is slightly redder than appropriate (see Fig. 
\ref{f:retmatch}), the bias is towards measurements which are too metal-rich. The reddening
estimates are dependent on both $(V-I)_g$ and $[$Fe$/$H$]$ (Sarajedini Eq. 3). The dependence on
$[$Fe$/$H$]$ is weak, so the primary error is introduced through $(V-I)_g$, which could be 
$\sim 0.02$ mag too red (see Section \ref{ss:cmdgood}). This would be transferred directly to
the $E(V-I)$ estimate, so it is possible these are too high by $\sim 0.02$. Nonetheless, the
consistency of our estimates with both literature measurements and the LMC distance scale lead
us to have confidence in the accuracy and validity of our results.

\subsection{Horizontal Branch Morphologies}
\label{ss:hbm}
Although we have images at only one epoch, and therefore no stellar variability information,
it is possible to calculate a quantitative measure of each cluster's HB morphology now that
accurate reddening values are known. As part of their study of $197$ RR Lyrae stars in four
globular clusters belonging to the Fornax dwarf galaxy, Mackey \& Gilmore \shortcite{fnx}
provided accurate measurements of the intrinsic $V-I$ colours of the red and blue edges of
the instability strip at the level of the horizontal branch. They found
$(V-I)_{{\rm BE}} = 0.28 \pm 0.02$ and $(V-I)_{{\rm RE}} = 0.59 \pm 0.02$. These values can
be used to count the number of blue HB stars, red HB stars, and stars on the instability strip.
The HB morphology is usually parametrized by the index $(B-R)/(B+V+R)$ of Lee, Demarque \& Zinn
\shortcite{ldz}, where $B$ is the number of BHB stars, $V$ the number of variable HB stars, and
$R$ the number of RHB stars. Because we have only single epoch observations it is possible that
some variables might lie outside the adopted instability strip edges (although this will not 
affect the morphology indices significantly, especially for NGC 1928 and 1939). In addition, because 
of the difficult field star subtractions necessary for NGC 1928 and NGC 1939, our number counts are 
commensurately uncertain. It is nonetheless worthwhile to make an attempt to calculate the HB
morphology for these clusters since no previous estimates exist.

\begin{table*}
\begin{minipage}{155mm}
\caption{Results of the CMD registration and relative age dating measurements (vertical technique).}
\begin{tabular}{@{}lccccccccccc}
\hline \hline
Cluster & \hspace{5mm} & $(V-I)_{{\rm TO}}$ & $V_{{\rm TO}}$ & \hspace{5mm} & $(V-I)_{0.05}$ & $V_{0.05}$ & \hspace{5mm} & $V_{{\rm HB}}$ & \hspace{5mm} & $\Delta V_{{\rm TO}}^{{\rm HB}}$ & $\Delta V_{0.05}^{{\rm HB}}$ \\
\hline
NGC 1928 & & $0.674$ & $22.851$ & & $0.724$ & $23.85$ & & $19.30$ & & $3.55$ & $4.55$ \\
NGC 1939 & & $0.660$ & $22.910$ & & $0.710$ & $23.69$ & & $19.30$ & & $3.61$ & $4.39$ \\
Reticulum & & $0.603$ & $22.627$ & & $0.653$ & $23.41$ & & $19.10$ & & $3.53$ & $4.31$ \\
\hline
M92 & & $0.558$ & $18.712$ & & $0.608$ & $19.49$ & & $15.18$ & & $3.53$ & $4.31$ \\
M3 & & $0.595$ & $19.173$ & & $0.645$ & $19.98$ & & $15.75$ & & $3.42$ & $4.23$ \\
M5 & & $0.629$ & $18.590$ & & $0.679$ & $19.44$ & & $15.18$ & & $3.41$ & $4.26$ \\
NGC 1466 & & $0.622$ & $22.860$ & & $0.672$ & $23.73$ & & $19.32$ & & $3.54$ & $4.41$ \\
NGC 2257 & & $0.612$ & $22.530$ & & $0.662$ & $23.40$ & & $19.10$ & & $3.43$ & $4.30$ \\
Hodge 11 & & $0.628$ & $22.812$ & & $0.678$ & $23.58$ & & $19.11$ & & $3.70$ & $4.47$ \\
\hline
\label{t:vages}
\end{tabular}
\end{minipage}
\end{table*}

\begin{table*}
\begin{minipage}{145mm}
\caption{Results of the relative age dating measurements (horizontal technique).}
\begin{tabular}{@{}llccllccccc}
\hline \hline
Cluster & $[$Fe$/$H$]$ & $\delta_{2.5}$ & \hspace{5mm} & Reference & $[$Fe$/$H$]$ & $\delta_{2.5}$ & \hspace{5mm} & $\delta(V-I)$ & \hspace{5mm} & $\Delta \tau$ \\
Name & & & & Cluster & & & & & & (Gyr) \\
\hline
NGC 1928 & $-1.27$ & $0.285$ & & M5 & $-1.27$ & $0.281$ & & $0.003$ & & $-0.2$ \\
NGC 1939 & $-2.10$ & $0.293$ & & M92 & $-2.28$ & $0.283$ & & $0.013$ & & $-1.3$ \\
 & & & & NGC 1466 & $-2.17$ & $0.289$ & & $0.006$ & & $-0.5$ \\
 & & & & & $(-1.85)$ & & & & & $(+0.6)$ \\
 & & & & Hodge 11 & $-2.06$ & $0.276$ & & $0.021$ & & $-0.9$ \\
Reticulum & $-1.66$ & $0.309$ & & M3 & $-1.54$ & $0.275$ & & $0.038$ & & $-1.4$ \\
 & & & & NGC 2257 & $-1.63$ & $0.290$ & & $0.023$ & & $-1.0$ \\
 & & & & & $(-1.85)$ & & & & & $(-1.8)$ \\
\hline
\label{t:hages}
\end{tabular}
\end{minipage}
\end{table*}

For NGC 1928 we counted $R = 2^{+1}_{-2}$, $V = 3^{+1}_{-3}$, and $B = 111^{+13}_{-11}$, including
the putative extended BHB ($\sim 11$ stars). For the more heavily populated NGC 1939 HB we counted 
$R = 3 \pm 3$, $V = 5^{+1}_{-5}$, and $B = 173^{+10}_{-12}$. These star counts result in a HB
index of $(B-R)/(B+V+R) = 0.94^{+0.06}_{-0.04}$ for both clusters, confirming their status
as having almost exclusively blue HB morphologies. Reticulum clearly has a more evenly spread
HB, and we count $R = 5^{+2}_{-4}$, $V = 18 \pm 3$, and $B = 5^{+0}_{-2}$, meaning a HB index
of $0.00 \pm 0.15$. This is entirely consistent with the result of Walker \shortcite{walker:ret}
who measured an index of $-0.04^{+0.00}_{-0.05}$. 

\subsection{Ages}
\label{ss:ages}
While it is quite evident from the CMDs presented in Fig. \ref{f:cmdsub} that the three clusters
from the present study exhibit all the photometric qualities of the oldest globular clusters,
it is useful to obtain some quantitative measure of their ages. The best way to achieve this
for the current sample is via differential comparison with clusters which have well established
ages. To this end, we employed cluster fiducials for the Galactic globular clusters M92, M3, and
M5 measured by Johnson \& Bolte \shortcite{johnson:98}, and for the LMC globular clusters
NGC 1466, NGC 2257, and Hodge 11 measured by Johnson et al. \shortcite{johnson:99}.

There is a large number of procedures available in the literature for the relative age dating of
globular clusters. Perhaps the two most widely used techniques are the so-called vertical and 
horizontal methods. The vertical method relies on the fact that the difference between the 
magnitude of the MSTO and the HB is age dependent, with older clusters generally having larger 
values of this parameter. Similarly, the horizontal method relies on the fact that the length of 
the SGB is shorter for older clusters, which thus have bluer RGBs. Both techniques have a
metallicity dependence which must be accounted for. Metallicities for the present LMC clusters
and the six reference clusters are listed in Table \ref{t:hages}. For M92, M3, and M5 these have
been obtained from the database of Harris \shortcite{harris}, while for NGC 1466, 2257, and Hodge 11
they are taken from various literature sources. Olszewski et al. measured spectroscopic abundances
for NGC 1466 and Hodge 11 ($[{\rm Fe}/{\rm H}] = -2.17$ and $-2.06$, respectively); however
Johnson et al. \shortcite{johnson:99} suggest a slightly more metal rich value for NGC 1466
($[{\rm Fe}/{\rm H}] = -1.85$). Johnson et al. also suggest $[{\rm Fe}/{\rm H}] = -1.85$ is
appropriate for NGC 2257 based on several previous estimates; however Dirsch et al. 
\shortcite{dirsch} obtained $[{\rm Fe}/{\rm H}] = -1.63$ based on a photometric study. 

For NGC 1928, NGC 1939, and Reticulum, we determined the colour and magnitude of the MSTO 
together with the MS reference point $0.05$ mag redder than the MSTO using exactly the same
technique as that described in Section \ref{ss:cmdgood}. The results may be found in Table 
\ref{t:vages}. These points were used for two
purposes -- measuring the vertical method age indicators, and registering CMDs for the horizontal
method (again, as described in Section \ref{ss:cmdgood}). For the reference clusters, we lacked
the full photometry sets, so could not apply exactly the same procedure. However, we found it
perfectly adequate to fit a quadratic function to the fiducial points around the MSTO
for these clusters. Johnson et al. \shortcite{johnson:99} provide measurements of $(V-I)_{{\rm TO}}$
and $V_{0.05}$ for NGC 1466, 2257, Hodge 11, M92, and M3. We find our calculations (again, see
Table \ref{t:vages}) to match their results very closely, which leads us to have confidence in
our procedure and our measurements for M5. We estimate that our determinations of $V_{{\rm TO}}$
and $V_{0.05}$ are accurate to $\pm 0.05$ mag, while those for $(V-I)_{{\rm TO}}$ and
$(V-I)_{0.05}$ are accurate to better than $\pm 0.01$ mag.

Our measured values for $\Delta V_{{\rm TO}}^{{\rm HB}}$, the difference in $V$ between the
level of the HB and the MSTO are listed in Table \ref{t:vages}. Both of these levels are quite
uncertain -- the HB because of some intrinsic width, as well as scatter due to stellar variability 
for Reticulum, and the significant weighting to the blue for NGC 1928 and 1939; and the MSTO because
the MS is vertical in the turn-off region. We estimate our measurement errors in 
$\Delta V_{{\rm TO}}^{{\rm HB}}$ to be approximately $\pm 0.1$ mag. The results in Table 
\ref{t:vages} show only a small dispersion among the clusters, even ignoring metallicity effects.
Rosenberg et al. \shortcite{rosenberg} used two sets of theoretically calculated isochrones to 
provide a calibration for age as a function of $\Delta V_{{\rm TO}}^{{\rm HB}}$ and metallicity. 
This appears in their Figure 3, which shows the significant majority of the $34$ Galactic globular 
clusters in their sample to lie within a narrow band of $\sim 2$ Gyr width about mean values of 
$14.3$ Gyr and $14.9$ Gyr for the two isochrone sets. Placing our clusters on this diagram (using
the metallicities listed in Table \ref{t:hages}) shows NGC 1928 and Reticulum to lie within
this band, along with all the reference clusters except Hodge 11. This cluster, along with NGC 1939
fall $\sim 2$ Gyr older than the upper limit of the Rosenberg et al. $2$ Gyr band.

In order to combat the uncertainty in $\Delta V_{{\rm TO}}^{{\rm HB}}$ introduced by measuring 
$V_{{\rm TO}}$, Buonanno et al. \shortcite{buonanno} introduced a calibration for a similar 
vertical parameter, $\Delta V_{0.05}^{{\rm HB}}$ -- the difference in $V$ between the level of the 
HB and $V_{0.05}$. Because the MS is sloped at $V_{0.05}$, this measurement is in theory more 
accurate (although formally, our errors are the same). We therefore calculated 
$\Delta V_{0.05}^{{\rm HB}}$ for each cluster for comparison with the Buonanno et al. calibration.
These measurements are also listed in Table \ref{t:vages}. The relevant calibration appears
in Figure 7(a)-(c) of Buonanno et al., for three isochrone sets. We note that this calibration is 
based on $(V\,,\,B-V)$ CMDs, however it should provide some indication of age homogeneity or 
otherwise in our cluster sample. Plotting the clusters on this Figure again shows them to lie within
a band of width $\sim 2$ Gyr, along with $14$ Galactic globular clusters from the Buonanno et al.
sample. The one outlier in our sample is NGC 1928, which lies $\sim 2$ Gyr older than the band.
For the three isochrone sets, the best fitting mean ages are $14$ Gyr, $12$ Gyr and $15$ Gyr,
respectively. 

Rosenberg et al. \shortcite{rosenberg} provide a calibration for a horizontal dating method in
addition to their vertical method calibration. The relevant parameter in this case is $\delta_{2.5}$,
the difference in $V-I$ between the MSTO and a point on the RGB $2.5$ mag brighter than the MSTO. 
Our measurements of this parameter for all the clusters are listed in 
Table \ref{t:hages}. Again, these results show a good deal of internal consistency, suggesting
small relative age differences. This is confirmed by consideration of Figure 4 in Rosenberg et al.
which shows their age calibration including metallicity effects, again for two isochrone sets.
As for the vertical age indicator, the $\delta_{2.5}$ values place all the present LMC and
reference clusters within the same $2$ Gyr-wide band. On this occasion, there are no significant
outliers.

\begin{figure*}
\begin{minipage}{175mm}
\begin{center}
\vspace{-5mm}
\includegraphics[width=165mm]{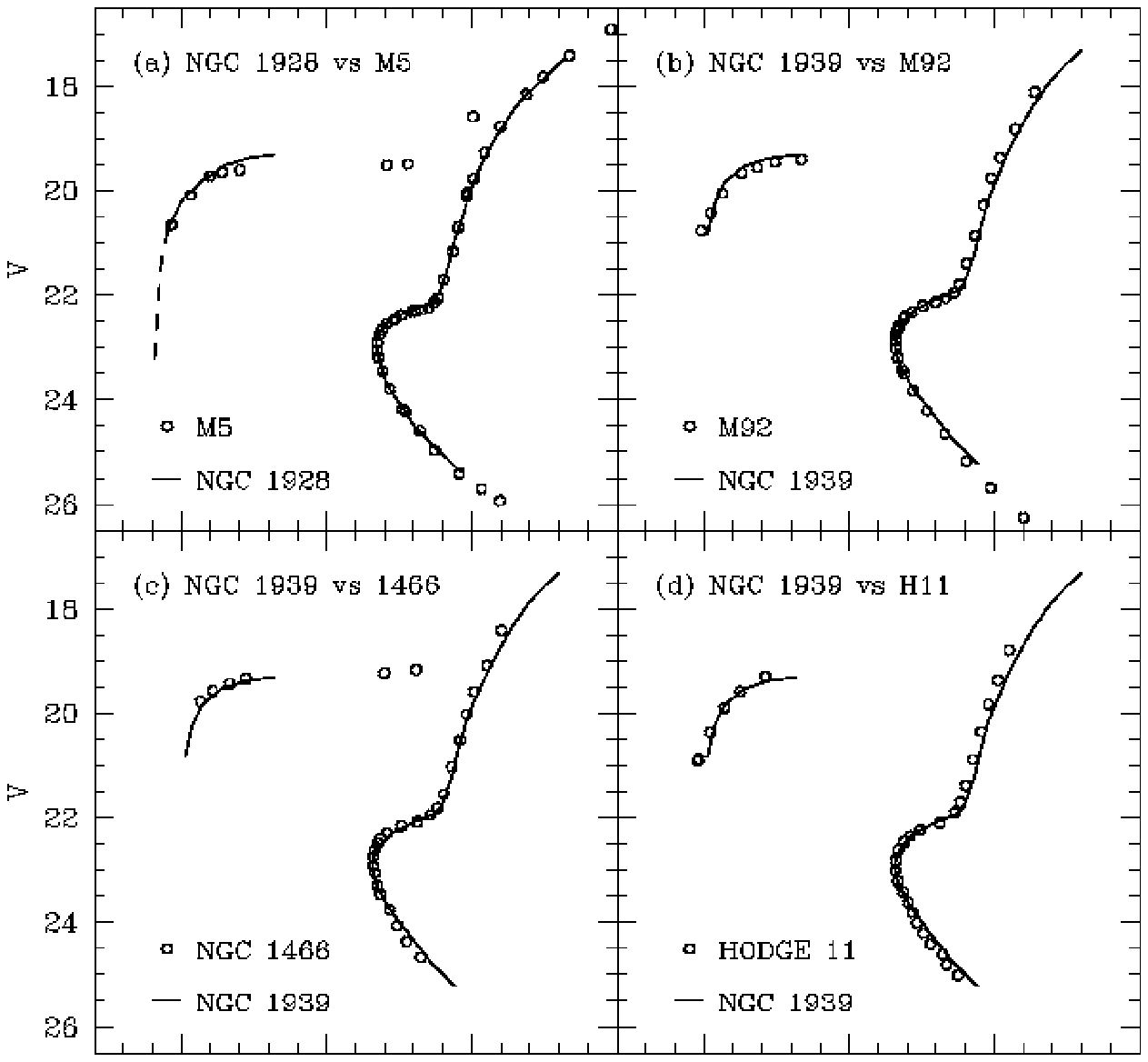} \\
\vspace{-19mm}
\includegraphics[width=165mm]{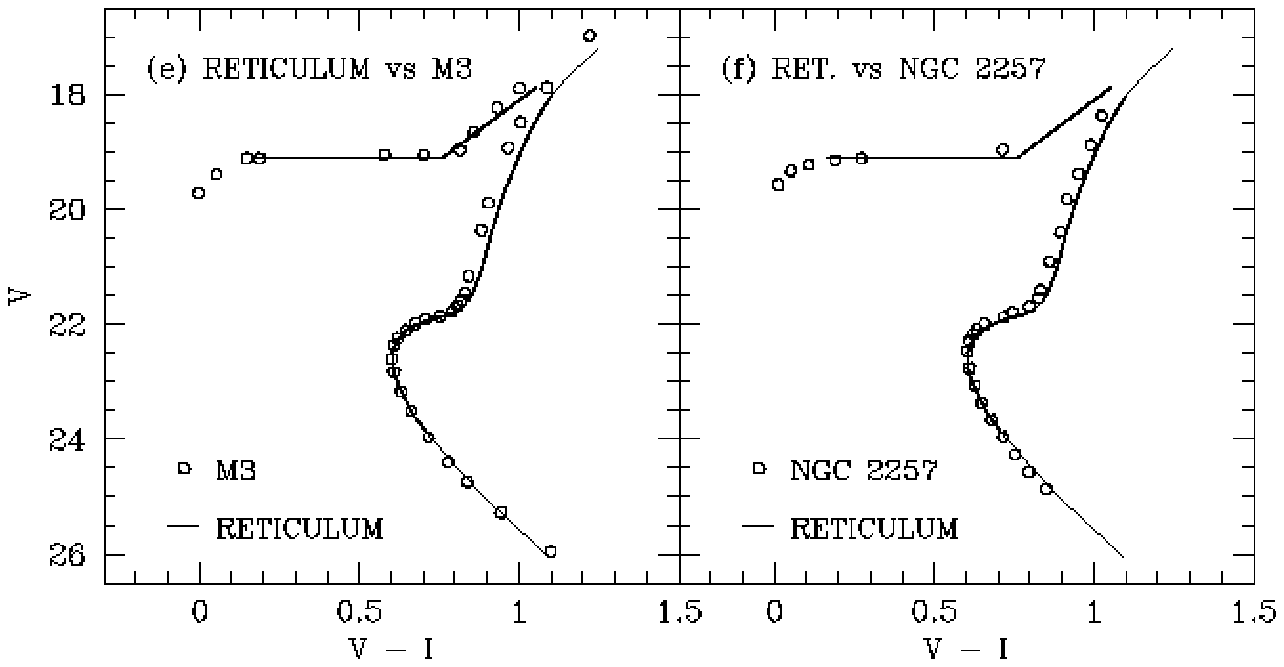}
\caption{Reference cluster fiducials (open circles) registered to the fiducials for NGC 1928, 1939, and Reticulum (solid lines) using $(V-I)_{{\rm TO}}$ and $V_{0.05}$. }
\label{f:fiducials}
\end{center}
\end{minipage}
\end{figure*}

Finally, we obtained a more quantitative measure of the relative cluster ages using the
horizontal method calibration of Johnson et al. \shortcite{johnson:99}. For this process, we first 
registered two cluster CMDs using $(V-I)_{{\rm TO}}$ and $V_{0.05}$, and then calculated the mean 
difference in colour between their RGBs from the base to the HB, $\delta(V-I)$. This was achieved 
using the procedure of Johnson et al. \shortcite{johnson:99}. A straight line was fit to the RGB 
section for the reference cluster, and then the weighted average of the difference in colour between
this line and the RGB stars for relevant the LMC cluster. Since $\delta(V-I)$ is sensitive to both 
age and metallicity, it was important to choose a reference cluster with similar metallicity to the 
LMC cluster under consideration. With the available clusters we had the following six matches: 
NGC 1928 and M5; NGC 1939 and M92, NGC 1466, and Hodge 11; Reticulum and M3, and NGC 2257. The 
results of the $\delta(V-I)$ calculations are listed in Table \ref{t:hages}, while Fig. 
\ref{f:fiducials} shows the reference cluster fiducials registered to the relevant LMC cluster
fiducials. It is immediately clear from these plots that there cannot be a large age difference
between the present three clusters and the oldest Galactic and LMC globular clusters. In order
to quantify this, we use the results of Johnson et al. \shortcite{johnson:99} who provide
relations between $\delta(V-I)$, age and metallicity using two theoretical isochrone sets.
The differences in the two calibrations are not large and we adopt a mean relation. First we
correct $\delta(V-I)$ between an LMC cluster and a reference cluster for metallicity effects
using
\begin{equation}
\delta(V-I) = -0.0745 (\Delta [{\rm Fe}/{\rm H}])
\end{equation}
and then estimate the age difference using
\begin{equation}
\delta(V-I) = -0.0205 (\Delta \tau)
\end{equation}
where $\tau$ represents age in Gyr. Strictly speaking this calibration is only valid over the
metallicity range $-2.26 < [{\rm Fe}/{\rm H}] < -1.66$; however it is unlikely that extrapolating
slightly to $[{\rm Fe}/{\rm H}] \sim -1.3$ for NGC 1928 and M5 will introduce a large error. 
The results of the calculations are listed in Table \ref{t:hages}. The two reference clusters
NGC 1466 and 2257 have differing metallicities listed in the literature, as discussed earlier.
Hence, these two clusters have a relative age range, as listed in Table \ref{t:hages}.

None of the present three LMC clusters is more than $\sim 1.5$ Gyr younger than any reference 
cluster. This offers quantitative evidence that NGC 1928, 1939, and Reticulum are true members of the
oldest population of LMC star clusters, and are coeval with the oldest Galactic globular clusters.
Supporting evidence is provided by the more qualitative measures offered by the vertical dating
technique -- the results of which are fully consistent with all the clusters being coeval to within
$\pm 1$ Gyr or so. We note that it is possible that our approximate transformation to Johnson $V$
and Cousins $I$ photometry has introduced some systematic error into the relative age measurements
-- after all, we noted earlier that the transformed RGBs are possibly too red by $\sim 0.02$ mag.
In this context, it is interesting to note that all the $\delta(V-I)$ values are positive,
indicating LMC cluster RGBs lying to the red of the reference cluster RGBs. This potential
systematic error does not affect our conclusion however, because redder RGBs indicate more
youthful clusters. Thus, the $\Delta \tau$ values listed in Table \ref{t:hages} represent the
maximum age differences implied by the $\delta (V-I)$ dating technique if the transformation is 
imperfect in the manner suspected -- in reality it is likely that the clusters are even closer in
age.

\subsection{NGC 1938}
\label{ss:1938}

\begin{figure*}
\begin{minipage}{175mm}
\begin{center}
\includegraphics[width=150mm]{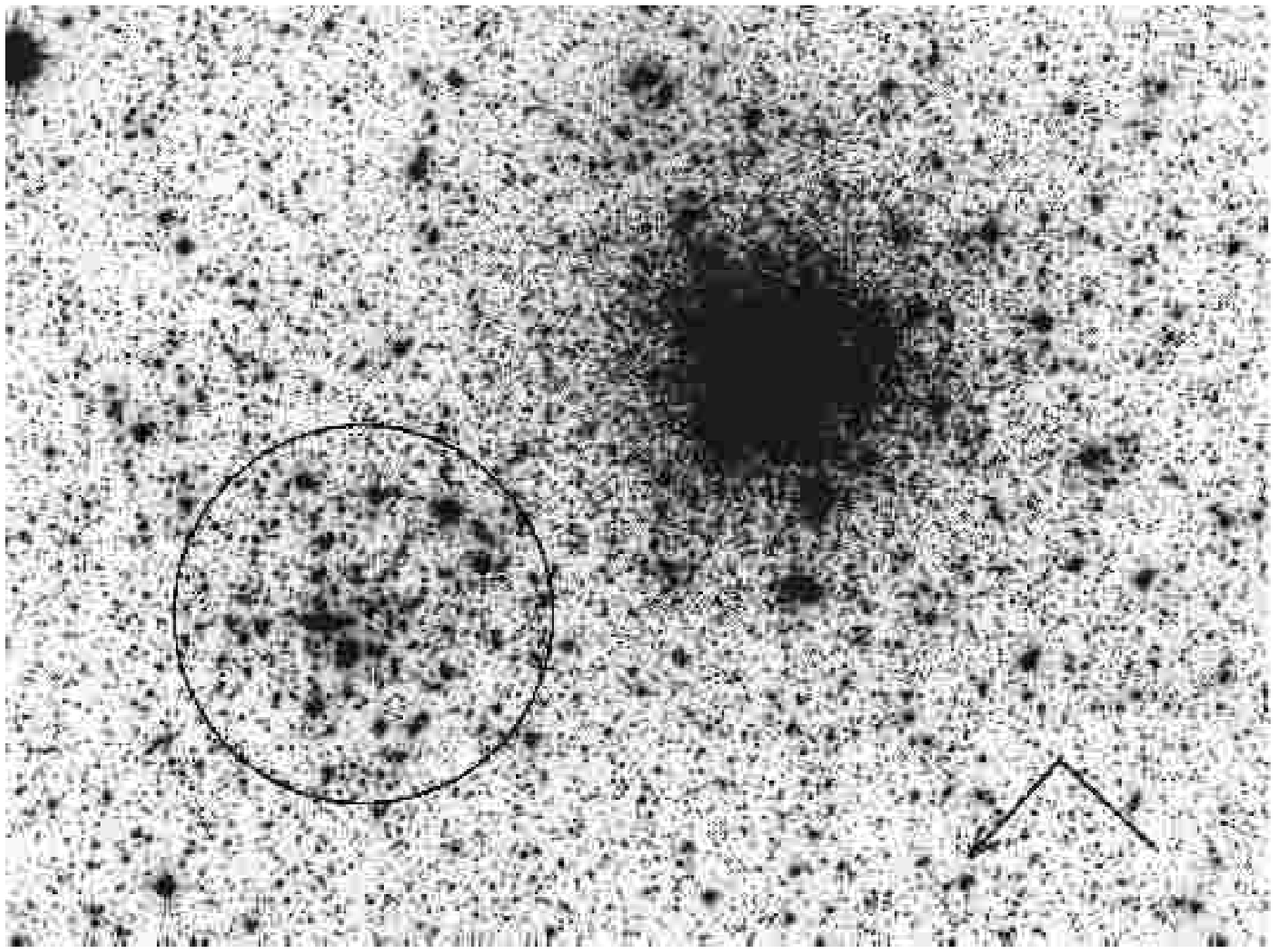}
\caption{ACS/WFC F555W image of the region surrounding NGC 1939 and NGC 1938 (circled). North and east are indicated (north is in the direction of the arrow). The NGC 1938 extraction radius is $\sim 12\arcsec$.}
\label{f:1938}
\end{center}
\end{minipage}
\end{figure*}

\begin{figure}
\includegraphics[width=0.5\textwidth]{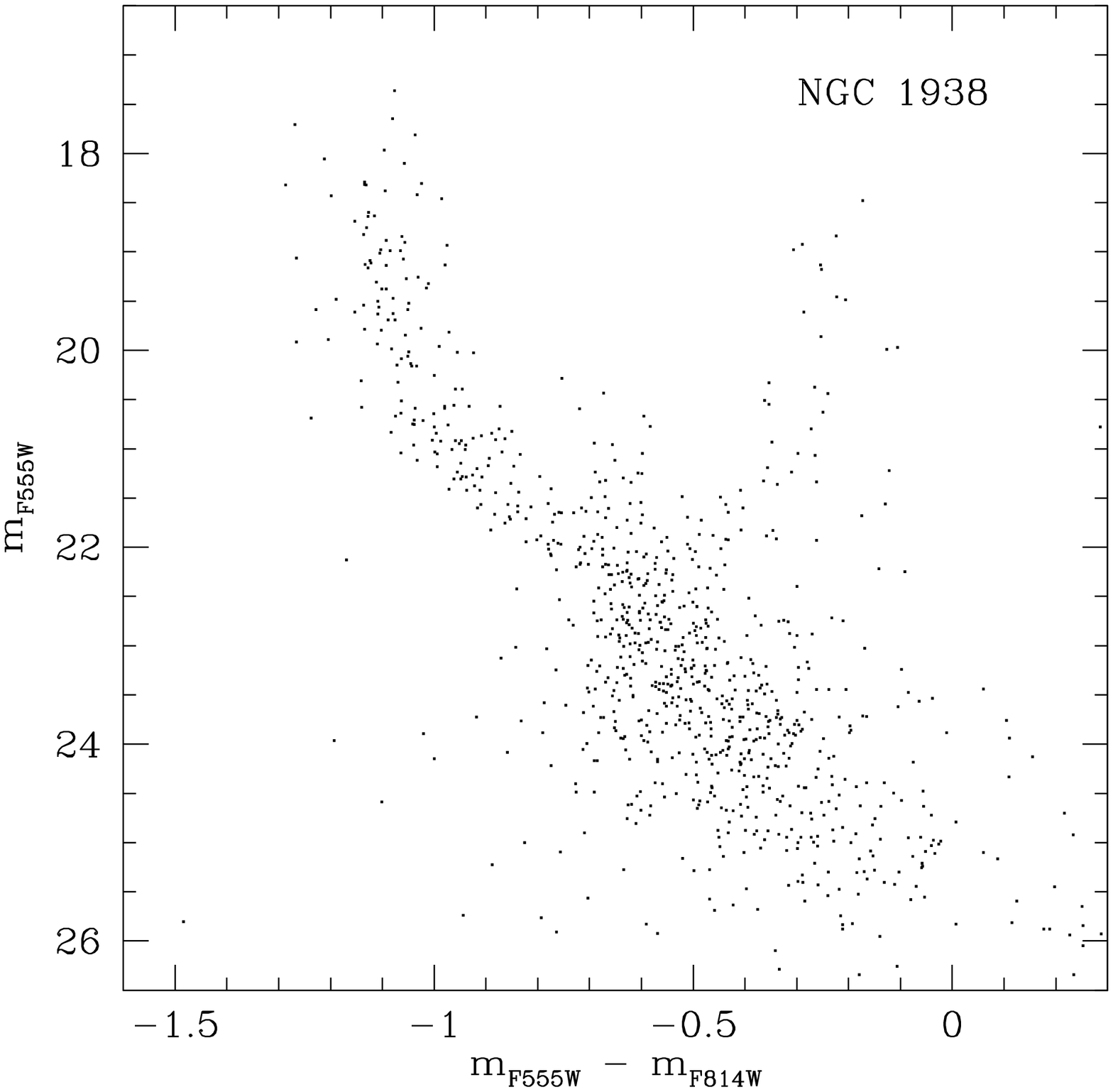}
\caption{CMD for all stars within the $12\arcsec$ extraction radius around NGC 1938. As previously, measurements are in the ACS/WFC STmag system.}
\label{f:cmd1938}
\end{figure}

It is worth plotting the CMD for the stars within $12\arcsec$ of the centre of NGC 1938 -- which
were removed from the calculations for NGC 1939 (see Section \ref{ss:fieldsub}) -- because we
could not find any previously published CMDs for this cluster. Fig. \ref{f:1938} shows a close-up
of the ACS/WFC F555W observation in the vicinity of NGC 1938 and 1939, with the extraction radius
shown. The resultant CMD for NGC 1938 is plotted in Fig. \ref{f:cmd1938}. No attempt has been
made to remove the field star contamination or the contamination from NGC 1939, the giant branch
of which is visible. Nonetheless, a narrow main sequence is clearly evident in this CMD. It is
not immediately obvious that this should be associated with NGC 1938 since it is clear from Fig.
\ref{f:cmdall} that it lies in the vicinity of the ``young'' main sequence of the field population.
To investigate this we calculated a crude fiducial for the sequence in Fig. \ref{f:cmd1938} above
$m_{{\rm F555W}} = 22$. We then counted all the stars less than $0.15$ mag perpendicular distance from
this fiducial in two bins -- an upper bin ($17 \le m_{{\rm F555W}} \le 20$) and a lower bin
($20 \le m_{{\rm F555W}} \le 22$). Next, two field extractions of equivalent area to the NGC 1938
extraction (i.e., with radius $12\arcsec$) were made away from both NGC 1938 and 1939 -- one at 
$40\arcsec$ to the north-west of NGC 1938, and one $1\arcmin$ to the north-east of NGC 1938. These
were far enough away not to be contaminated by NGC 1938 or NGC 1939, and close enough not to be
affected by serious differential reddening. We then repeated our star counts in these two extractions.
In the on-cluster extraction we counted $64$ stars in our upper bin and $107$ in our lower bin.
For the two off-cluster extractions we counted $5$ and $8$ stars respectively for the upper bin,
and $31$ and $33$ stars respectively for the lower bin. The star counts are clearly significantly 
enhanced for the on-cluster field, especially for the upper main sequence region. This allows us
to confidently associate the sequence observed in Fig. \ref{f:cmd1938} with NGC 1938 and not the
general field population.

No turn-off is observed on this sequence to $m_{{\rm F555W}} \approx V \approx 18$. We use this fact to 
place an upper limit on the age of NGC 1938. Kerber et al. \shortcite{kerber} provide a CMD for 
the rich LMC cluster NGC 1831 from WFPC2 observations, which shows the turn-off to be just below 
$V \approx 18$. Their age estimate for this cluster is $\tau \approx 500$ Myr. Elson \& Fall 
\shortcite{elson:88} provide an age calibration for a large number of LMC clusters. Their age 
estimate for NGC 1831, corrected to a distance modulus of $18.5$ is $\sim 300$ Myr. Together 
these age estimates provide us with an approximate upper limit for the age of NGC 1938 of $\sim 400$
Myr. This result is perfectly consistent with the integrated $UBV$ photometry of 
Bica et al. \shortcite{bica}, which places NGC 1938 in SWB class IVA (where the SWB class refers 
to the classification of Searle, Wilkinson \& Bagnuolo \shortcite{swb}). This class brackets the 
age range $200$--$400$ Myr. Because of the very large age difference between NGC 1938 and 1939, 
it seems unlikely that these two clusters are physically related -- rather, their apparent close 
proximity is merely a projection effect.

\section{Summary}
\label{s:summary}
We have used ACS/WFC snapshot observations to obtain colour-magnitude diagrams for the LMC 
clusters NGC 1928, 1939 and Reticulum. This is the first time that CMDs for NGC 1928 and 1939
have been published. These two CMDs suffer from very dense field star contamination requiring
a thorough subtraction algorithm. Using the final CMDs we obtained photometric reddening and 
metallicity measurements for all three clusters. NGC 1939 is one of the most metal-poor 
LMC bar clusters, with $[{\rm Fe}/{\rm H}] = -2.10 \pm 0.19$, while NGC 1928 is significantly
more metal-rich, with $[{\rm Fe}/{\rm H}] = -1.27 \pm 0.14$. Reticulum is of a more intermediate
abundance, with $[{\rm Fe}/{\rm H}] = -1.66 \pm 0.12$. This measurement matches well the 
previous estimates for this cluster.

All three clusters possess CMDs with features characteristic of the oldest Galactic globular 
clusters -- main sequence turn-offs at $V\sim 23$, and well populated horizontal branches.
Both NGC 1928 and 1939 possess very blue HB morphologies, with little or no population stretching
to the red of the blue edge of the instability strip. In contrast, Reticulum has a HB
populated right across the instability region. To quantify the ages of the three clusters 
we employed a variety of differential dating techniques, comparing their CMDs to a set of three of
the oldest Galactic globular clusters, and three of the oldest LMC globular clusters. We conclude
that the entire set of clusters is coeval to within approximately $2$ Gyr. This work firmly 
establishes NGC 1928 and 1939 as members of the LMC globular cluster population, confirming the
conclusion obtained by Dutra et al. \shortcite{dutra} from integrated spectroscopy. The
LMC globular cluster census therefore now numbers $15$, in two distinct groups -- the outer
clusters NGC 1466, 1841, 2210, 2257, Hodge 11, and Reticulum; and the inner (bar) clusters
NGC 1754, 1786, 1835, 1898, 1916, 1928, 1939, 2005, and 2019. The only other LMC cluster
older than the lower end of the age gap is ESO121-SC03, at $\sim 9$ Gyr.

\section*{Acknowledgements}
This paper is based on observations made with the NASA/ESA 
{\em Hubble Space Telescope}, obtained at the Space
Telescope Science Institute, which is operated by the Association
of Universities for Research in Astronomy, Inc., under NASA
contract NAS 5-26555. These observations are associated with
program 9891. ADM is grateful for financial support from PPARC
in the form of a Postdoctoral Fellowship.

% End of paper here

%%%%%%%%%%%%%%%%%%%%%%%%%%%%%%%%%%%%

\bsp % ``This paper has been produced using the ...''

\label{lastpage}

\end{document}